\documentclass[twocolumn]{IEEEtran}
\IEEEoverridecommandlockouts

\usepackage{amsmath, amssymb, latexsym, amsthm, graphicx, epsfig, subfigure, cite, algorithm, algpseudocode, pifont, xcolor, scalerel, enumerate, multicol, lipsum, cuted, stfloats, booktabs}
\usepackage{array}
\usepackage{diagbox}
\usepackage{calc}
\usepackage{makecell}
\usepackage{cases}
\newcolumntype{P}[1]{>{\centering\arraybackslash}p{#1}}

\allowdisplaybreaks

\makeatletter
\newcommand{\doublewidetilde}[1]{{
		\mathpalette\double@widetilde{#1}}}
\newcommand{\double@widetilde}[2]{
	\sbox\z@{$\m@th#1\widetilde{#2}$}
	\ht\z@=.5\ht\z@
	\widetilde{\box\z@}}
\makeatother

\usepackage{accents}

\newtheorem{lemma}{Lemma}

\newtheorem{proposition}{Proposition}

\newtheorem{remark}{Remark}

\begin{document}

\title{\huge Graph Neural Network   based Active and Passive Beamforming  for Distributed STAR-RIS-Assisted Multi-User MISO Systems 
}

\author{Ha An Le, Trinh Van Chien, \textit{Member}, \textit{IEEE}, and Wan Choi, \textit{Fellow}, \textit{IEEE} 
\thanks{Ha An Le and Wan Choi are with the Department of Electrical and Computer Engineering, Seoul National University, Seoul, Korea}
\thanks{Trinh Van Chien is with the School of Information and Communications Technology (SoICT), Hanoi University of Science and Technology, Hanoi 100000, Vietnam.}
  }

\maketitle

\begin{abstract}
This paper investigates a joint active and passive beamforming design for distributed simultaneous transmitting and reflecting (STAR) reconfigurable intelligent surface (RIS) assisted multi-user (MU)- mutiple input single output (MISO) systems, where the energy splitting (ES) mode is considered for the STAR-RIS. We aim to design the active beamforming vectors at the base station (BS) and the passive beamforming at the STAR-RIS to maximize the user sum rate under transmitting power constraints. The formulated problem is non-convex and nontrivial to obtain the global optimum due to the coupling between active beamforming vectors and STAR-RIS phase shifts. To efficiently solve the problem, we propose a novel graph neural network (GNN)-based framework. Specifically, we first model the interactions among users and network entities are  using a heterogeneous graph representation. A heterogeneous graph neural network (HGNN) implementation is then introduced to directly optimizes beamforming vectors and STAR-RIS coefficients with the system objective. Numerical results show that the proposed approach yields efficient performance compared to the previous benchmarks. Furthermore, the proposed GNN is scalable with various system configurations.
\end{abstract}


\section{Introduction}
Beyond fifth-generation (B5G) and sixth-generation (6G) wireless communication networks are expected to cope with the explosive increase in the number of wireless devices with a focus on spectral and energy efficiency\cite{WSaad2019-6GVision}. In this light, RISs have emerged as a promising technology capable of significantly enhancing the sum rate and energy efficiency of wireless networks \cite{Ebasar2019-RISintro}.  RIS is a flat meta-surface containing a number of inexpensive passive reflecting components, which can be adjusted through a controller to smartly manage the propagation of the incident signals with low power consumption. Moreover, RIS deployment allows for the establishment of a connection between the BS and user equipment (UE), especially in scenarios where they are situated in areas with no service or where direct links are obstructed. Nevertheless,  typical RISs are only capable of reflecting incident signals, hence only users located in the $180^o$ half-plane can be supported by the RISs. Consequently, the positioning of both the BS and users is constrained to be on the same side as the RISs, which restricts their deployment. As a remedy, the STAR-RIS emerges as a promising technology that overcomes the constraints of conventional RIS setups as it can extend the coverage from half-space to a complete $360^o$ space\cite{Mu2022-STARRIS}. Specifically, the STAR-RIS divides the three-dimensional (3D) space into two distinct regions, i.e., the transmission region ($\mathcal{T}$ region) and the reflection region ($\mathcal{R}$ region). Therefore, compared to conventional RISs, STAR-RISs provide new degrees-of-freedom (DoFs) that enhance the system performance \cite{Mu2022-STARRIS}.

Due to its significant potential, researchers in both industry and academia have directed considerable attention toward STAR-RIS and its variations. Many works have exploited the STAR-RIS in various system settings.
In \cite{Mu2022-STARRIS}, the authors examined a MISO system aided by STAR-RIS, and focused on a problem aiming at minimizing power consumption while considering active and passive beamforming. 
In \cite{Wang2023-STARRISNOMA},
the authors delved into a problem of optimizing both  active beamforming vectors and STAR-RIS phase shifts to enhance the energy efficiency and overall sum rate of NOMA systems. An iteration based semidefinite relaxation (SDR) scheme was proposed to tackle the non-convexity. 
The authors in \cite{Wu2021-STARRISCoverage}, examined the coverage characterization of a two-user system aided by STAR-RIS with a joint optimization of power allocation at the access point and the STAR-RIS coefficients. The non-convex decoding order constraint in the problem was re-transformed into a convex one by applying KKT conditions.
Nevertheless, most of the current research has focused on single STAR-RIS-assisted wireless systems, while the benefits of deploying multiple intelligent reflecting surfaces have been investigated in  \cite{Papaza2021-MulRISCoverage,Wang2020-MulRIS,Yang2022-MulRISEE,Jinkyu2024-RIS}. As opposed to the single RIS scenario, the distributed employment has revealed its potential for enhancing  coverage, signal power, and system energy efficiency. Therefore, it is crucial to extend the study of STAR-RIS-assisted systems to encompass the distributed scenario, which is the main focus of our paper.

A key challenge in the STAR-RIS system is computational cost. Unlike conventional RIS systems, transmission and reflection  elements in the STAR-RIS are coupled together, which further increase the resource allocation complexity \cite{Mu2022-STARRIS}. In this light, deep learning (DL) has stood out as a cost-effective solution for the RIS assisted system optimization. It has been illustrated that traditional DL models, such as fully connected neural network (FCNN) and CNN, can greatly reduce system computational cost while maintaining a comparable performance in various applications \cite{Gao2020-RISFCNN,HaAn2023-DoubleRIS,WXU2022-RISDQNN,Zhong2022-STARRISHybridDRL}.
Nevertheless, these DL models lack the versatility to generalize across various network sizes, such as differing numbers of users and RIS elements. This limitation stems from the fixed output/input dimensions inherent in these models; thus, an FCNN/CNN trained with a specific configuration cannot be seamlessly applied to others. Consequently, the practical utility of DL models is significantly hampered, especially in dynamic system configurations. Moreover, conventional DL models, e.g. FCNN/CNN, optimize the network by learning the mapping between the input and optimal output rather than the underlying topology of the network. Therefore, in a large system, the performance of these models degrades significantly \cite{Shen2023-GNNWireless}. 

In this light, graph neural networks (GNNs) stand out as a promising solution. GNN consists of vertices and edges that are meticulously engineered to ensure their feature dimensions remain invariant to different network sizes. This approach preserves both permutation invariance (PI) and permutation equivariance (PE) within the model \cite{Shen2023-GNNWireless}. In other words, GNNs are able to learn the underlying interaction between network entities and generalize well with the varying numbers of order of them. 

With a promising potential, GNNs have been extended to solve many scalable wireless communication systems, such as power allocation, beamforming design, and so on \cite{Eisen2020-GNNPowercontrol,Shen2021-GNNPA,Chowdhury2021-GNNWMMSE,Jiang2021-GNNRIS,Lyu2024-RISGNN,Wang2023-RISGNNFL,Liu2023-RISGNN}. 
Regarding RIS-aided networks, homogeneous GNN networks are proposed in \cite{Jiang2021-GNNRIS,Lyu2024-RISGNN} to jointly optimize transmit precoding vector at the BS and RIS phase-shift from the received pilot signal. Similarly, the authors in \cite{Wang2023-RISGNNFL} utilized a homogeneous GNN model to optimize the RIS-assisted over-the-air federated learning (FL) network. In these works, each edge device/user is modeled as a vertex, and the RIS vertex is initialized by averaging the features of all user vertices. These works demonstrate that the proposed GNN design can generalize well across different numbers of users. Specifically, while the model is trained under a specific system configuration, it achieves adequate performance when tested on unseen configurations with varying numbers of users. However, in these designs, the RIS coefficient matrix is predicted directly from the RIS vertex's feature, which ties the {vertex's feature dimension to the number of RIS elements. Consequently, the model's scalability to different RIS sizes is limited. In addition, these designs focus on single RIS setting and are not well-suited for the multiple RIS scenarios due to several reasons. First, the RIS vertex's feature  in these models is formed by averaging the features of user vertices, which might work for single-RIS systems but fails to capture the unique characteristics of each RIS and their specific connections to users. Second, since the  same feature is shared among RIS vertices, which is independent of their ordering, permuting RIS vertices does not result in a corresponding permutation of the output associated to RIS vertices. Consequently, the PE property with respect to RISs is not preserved, leading to potential performance degradation and limited generalization\cite{Guo2022-HetnetBF}. {Therefore, it is challenging to extend these designs to the multi-RISs scenarios.

For multi-RIS systems, recognizing the limitations of DNNs in capturing the underlying structure of wireless networks, the authors of \cite{Liu2023-RISGNN} proposed a heterogeneous GNN model for joint beamforming design in a multi-RIS-aided mmWave system. In this design, each user, RIS, and BS are treated as graph vertices, with their features derived from channel information. By more effectively capturing interactions among system entities, the proposed GNN model significantly outperforms conventional DNN approaches. Nevertheless, the proposed design for multi-RIS systems also encounters limitations. First,  user vertex's feature is initialized by concatenating cascaded channel associated with RISs.  While this design captures sufficient channel information, its feature dimension expands with the number of RISs since user vertex's feature is constructed by concatenating BS-RIS-User channel corresponding to all RISs. Second, the feature dimension of the BS vertex is defined by the transmit beamforming matrix, while the feature dimension of the RIS vertex is characterized by the reflecting coefficient matrix for each RIS. Consequently, the feature dimension of the BS vertex depends on the number of users and transmit antennas, whereas the feature dimension of the RIS vertex is determined by the size of the RIS. As a result, the model proposed in \cite{Liu2023-RISGNN} lacks the ability to generalize across different system configurations. Furthermore, while existing works have proposed GNN models for RIS system optimization, they often overlook the specific PE properties of the system. Although such models may incidentally inherit certain PE properties, it is crucial to ensure that the properties of the designed GNN model are intentionally aligned with the PE properties of the optimization policy. Such alignment improves the system's generalization across varying network configurations and sizes \cite{Shen2021-GNNPA,Guo2022-HetnetBF}.

Motivated by the potential of multiple STAR-RISs and the existing gap in the literature on distributed STAR-RIS systems, this paper investigates the joint active and passive beamforming design in a distributed STAR-RIS-aided multi-user MISO system. We first derive an analytical solution to the design problem. To address the non-convex nature of this problem, we adopt an alternating optimization (AO)-based successive convex approximation (SCA) framework, which guarantees convergence to a KKT solution. However, since the computational complexity of this analytical approach is too high, the scalability of the AO-based framework is constrained. 

To achieve a scalable design, we then propose a Beamforming Heterogeneous Graph Neural Network (BHGNN) framework specifically tailored to the distributed STAR-RIS system. 
In particular, we leverage a graphical representation to model the interconnections between the entities in the STAR-RIS systems. Unlike existing works, the proposed graphical model is designed to be independent of system settings, such as the number of STAR-RISs, users, and the size of the STAR-RIS. Specifically, we treat each STAR-RIS element and user as a vertex, eliminating the dependency of the vertex's feature dimension on the size of the STAR-RIS. Additionally, the interconnections between these vertices are defined by the corresponding equivalent wireless channels, which remain invariant to the number of STAR-RISs and users. Consequently, this framework generalizes across various system settings.
Additionally, we introduce a heterogeneous graph message-passing (HGMP) procedure to facilitate information exchange among vertices, which is rigorously proven to preserve the permutation equivariance (PE) property of the optimal beamforming policy. Our results show that the proposed BHGNN achieves performance close to that of the AO method while significantly reducing computational complexity. 
Our primary contributions are outlined as follows:
\begin{itemize}
    \item First, we formulate a sum-rate optimization problem for distributed STAR-RIS-assisted MU-MISO systems. To derive an analytical solution, we decompose the original non-convex problem into sub-problems, which are then approximated by a convex problem and solved iteratively. Although the AO-based SCA framework is proven to converge to the KKT solution, its prohibitively high complexity limits scalability, underscoring the need for a more practical and scalable design.
    \item Second, we investigate the PE property of the optimal beamforming policy in the considered system and propose a heterogeneous graph representation to model the considered wireless networks.  In our graph representation, each STAR-RIS element and user is treated as a vertex, rather than the entire STAR-RIS as a single vertex, enabling the scalability of the network to various STAR-RIS sizes. The features of the corresponding vertices and edges are carefully designed to ensure invariance to the number of vertices, enabling scalability with various system configurations. Consequently, we propose a HGMP algorithm dedicated to beamforming tasks which facilitates the information exchange through the entire graph. We then prove that the PE property is well-preserved in the proposed HGMP.
    \item Third, we present an effective implementation of the BHGNN model executing the proposed message-passing algorithm.  Specifically, each function within the HGMP algorithm is approximated by an FCNN model, which is reused at all vertices with the same type, i.e. STAR-RIS and user vertex. The forward propagation of the designed BHGNN model is executed aligned with the HGMP algorithm and the model is then trained to optimize the system sum rate. Given that the dimensions of vertices and edges remain unaffected by the system's configuration, our designed BHGNN demonstrates strong generalization capabilities across various system settings, rendering it suitable for dynamic networks. 
    \item Finally, we produce extensive simulations to validate the efficacy of the proposed methodology. The numerical results indicate that the BHGNN can attain performance levels close to those of the AO-based approach with much lower computational complexity. Additionally, the HGNN model exhibits robust generalization across varying numbers of users, STAR-RIS elements, and user distributions. 
\end{itemize}
The rest of this paper is organized as follows: Section~\ref{sec:Sysmodel} introduces the system model and outlines the formulation of the problem aiming to maximize user sum rates an AO-based solution addressing the formulated problem. Section~\ref{Sec:BGNN} introduces a heterogeneous graph representation for the system and proposes a GNN-based solution to jointly optimize beamforming vectors  and STAR-RIS phase shifts, maximizing the system sum rate. The numerical results and discussions are detailed in Section~\ref{sec:Results}, and Section~\ref{sec:Conclusion} offers the primary conclusions drawn from this paper.

\textit{Notation}: Matrices are presented by bold capital letters, and lower bold letters denote vectors. The regular transpose and Hermitian transpose of a matrix $\mathbf{A}$ are denoted by $\mathbf{A}^T$ and $\mathbf{A}^H$, respectively. The trace of a square matrix $\mathbf{A}$ is denoted by $\mathrm{tr}(\mathbf{A})$, while its inverse is represented as $\mathbf{A}^{-1}$. Furthermore, $\mathrm{Re}(.)$ and $\mathrm{Im}(.)$ are employed to represent the real and imaginary parts, respectively, of the given argument. 
$\| \cdot \|$  represents the Euclidean norm, and $\mathcal{CN}(\cdot, \cdot)$ is a circularly symmetric Gaussian distribution. $\nabla f(\cdot)$ denotes the derivative of function $f(\cdot)$. Finally, $\mathcal{O}(\cdot)$ is the big-$\mathcal{O}$ notation.
  
\section{System model and problem formulation} \label{sec:Sysmodel}
This section describe the distributed STAR-RIS system under study in the paper. We formulate a problem aimed at maximizing the sum rate while adhering to constrained power limits, specifically addressing the joint design of active and passive beamforming.

\subsection{System model}
We investigate a distributed MISO system aided by STAR-RIS, where a BS with $N_t$ antennas serves $K$ individual users, each equipped with a single antenna, positioned on either the transmission side $(\mathcal{T})$ or the reflection side $(\mathcal{R})$. To enhance the system spectral efficiency, $L$ STAR-RISs are deployed to support the transmission, each of which contains $M$ elements. 

In this paper, we make the assumption that the STAR-RISs operate in the energy splitting (ES) mode, which entails partitioning the power of the incoming signal into transmitted and reflected signal energies. Consequently, the matrices representing the transmission and reflection coefficients of the $l$-th STAR-RIS are expressed as
\begin{equation}
\pmb{\Phi}_l^{\chi} = \mathrm{diag}\left(\sqrt{\beta_{l1}^\chi}e^{\theta_{l1}^{\chi}},\cdots,\sqrt{\beta_{lM}^\chi}e^{\theta_{lM}^{\chi}}\right),
\end{equation}
where $\chi \in \{\mathcal{T},\mathcal{R}\}$. For brevity, we represent $\pmb{\beta}_{l}^\chi = [\sqrt{\beta_{l1}^\chi},\cdots,\sqrt{\beta_{lM}^\chi}]^T$ and $\mathbf{\Theta}_l^\chi = [e^{j\theta_{l1}^\chi},\cdots,e^{j\theta_{lM}^\chi}]^T$ as the amplitude and phase shift vectors, respectively, of the $l$-th STAR-RIS in the transmitter and receiver regions. Additionally, this study assumes that direct links between the base station (BS) and users are obstructed by obstacles. Among the $K$ users, we consider the initial $K_0$ users positioned in the $\mathcal{T}$ region, while the rest $(K- K_0)$ users are situated in the $\mathcal{R}$ region. Specifically,  $\chi = \mathcal{T}$ if $k \in {1,2,\cdots, K_0}$, otherwise $\chi = \mathcal{R}$. The received signal at the $k$-th user is expressed as
\begin{equation}\label{eq:RecievedSignal}
\begin{split}
      y_k &= \sum_{l=1}^{L}\mathbf{h}_{kl}^H\mathrm{diag}(\pmb{\beta}_l^\chi)\mathrm{diag}(\mathbf{\Theta}_l^\chi)\mathbf{G}_l\mathbf{x} + n_k, \\     &=\sum_{l=1}^{L}\mathbf{h}_{kl}^H\pmb{\Phi}_l^{\chi}\mathbf{G}_l\mathbf{x} + n_k,
\end{split}
\end{equation}
where $\mathbf{x} = \sum_{k=1}^{K}\mathbf{w}_ks_k$ represents the transmitted symbols, with $\mathbf{w}_k$ represents the precoding vector associate with the $k$-th user, and  $n_k \sim \mathcal{CN}(0,\sigma^2)$ denotes additive white Gaussian noise. Moreover, $\mathbf{h}_{kl} \in \mathbb{C}^{M}$ represents the channel from the $l$-th RIS to the $k$-th user, and $\mathbf{G}_l \in \mathbb{C}^{M \times N_t}$ signifies the channel from the BS to the $l$-th RIS. The signal-to-interference-and-noise ratio (SINR) at the $k$-th user is represented as
\begin{equation} \label{eq:SINR}
    \gamma_k = \frac{\left|\sum_{l=1}^{L}\mathbf{h}_{kl}^H\pmb{\Phi}_l^{\chi}\mathbf{G}_l\mathbf{w}_k\right|^2}{\sum_{j \in \mathcal{K}, j\neq k}\left| \sum_{l=1}^{L}\mathbf{h}_{kl}^H\pmb{\Phi}_l^{\chi}\mathbf{G}_l\mathbf{w}_{j}\right|^2+\sigma^2},
\end{equation}
where $\mathcal{K} = \{1,2,\cdots,K\}$ is the set of user. 
\begin{figure} [t]
    \centering
    \includegraphics[trim=0cm 0cm 0cm 0cm, clip=true, width=3.2in]{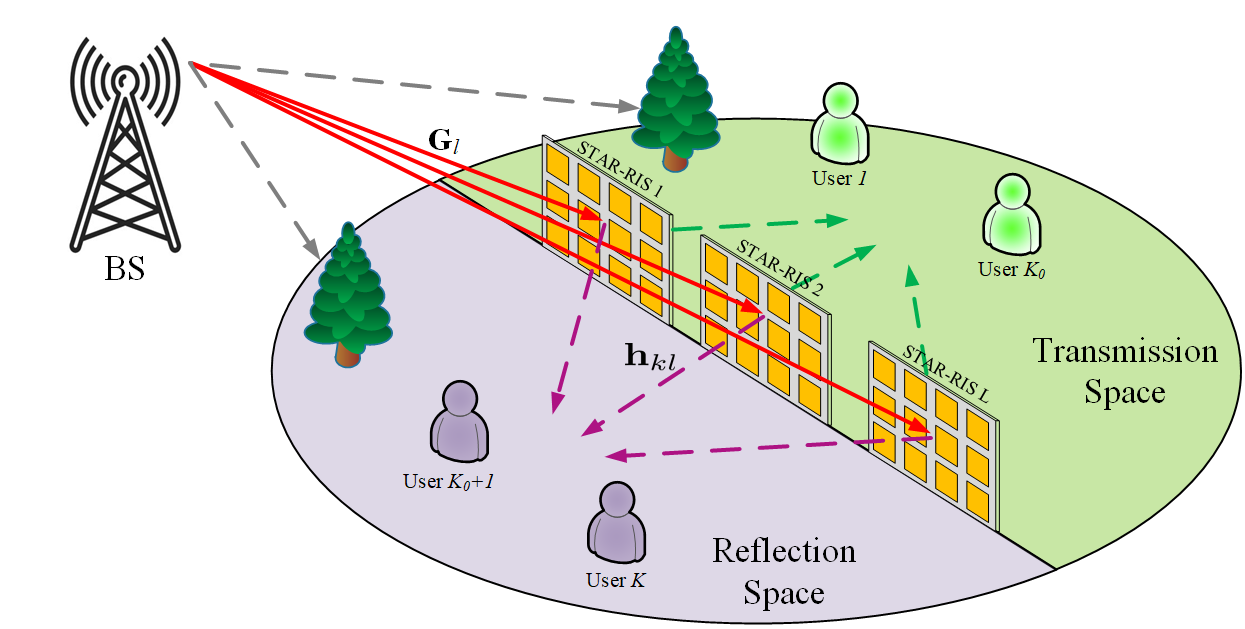}
    \caption{The considered distributed STAR-RIS system model.}
    \label{fig:SystemModel}
    \vspace{-0.3cm}
\end{figure}

\subsection{Problem Formulation}
Our objective is to maximize the system sum rate of all users while adhering to the transmit power constraint by jointly optimizing the precoding vector $\mathbf{w}_k$, the STAR-RIS phase shift $\mathbf{\Theta}^\chi$, and the amplitude coefficients $\pmb{\beta}_{l}^\chi$. Mathematically, the optimization problem is formulated as
\begin{subequations}\label{eq:ProbP1}
\begin{align}
    &\underset{\{\mathbf{W},\mathbf{\Theta}^\chi,\pmb{\beta}^\chi\}·}{\mathrm{maximize}} &&\sum_{k \in \mathcal{K}} \log_2(1+\gamma_k) \label{eq:sumrate}\\
    & \mbox{subject to} &&\mathbf{w}^H\mathbf{w} \leq P_{\max} \label{powerConstraint}, \\
    & && \theta_{lm}^\chi \in [0,2\pi], \quad \forall l \in \mathcal{L}, m \in \mathcal{M},\label{phaseConstraint}\\
    & && 0 \leq \beta_{lm}^\chi \leq 1 \quad \forall  l \in \mathcal{L}, m \in \mathcal{M},\label{constraint3} \\ 
    & &&  (\beta_{lm}^T)^2 + (\beta_{lm}^R)^2 \leq 1, \forall  l \in \mathcal{L}, m \in \mathcal{M}, \label{constraint4}
\end{align}    
\end{subequations}
where $\mathcal{M} = \{1,2,\cdots,M\}$ is the set of STAR-RIS phase shift elements, $\mathcal{L} = \{1,2,\cdots,L\}$ is the set for STAR-RISs. The stacked beamforming vector of all $K$ user is $\mathbf{w} = [\mathbf{w}_1^T,\mathbf{w}_2^T,\cdots,\mathbf{w}_K^T]^T \in \mathbb{C}^{N_tK}$. Similarly, the stacked phase shift matrix of all the $L$ STAR-RISs is  $\mathbf{\Theta}^\chi = [(\mathbf{\Theta}_1^\chi)^T,\cdots(\mathbf{\Theta}_L^\chi)^T]^T$, $\pmb{\beta}^\chi = [(\pmb{\beta}_1^\chi)^T, \cdots,(\pmb{\beta}_L^\chi)^T]^T$. In \eqref{eq:ProbP1},  $P_{\mathrm{max}}$ is the transmit power budget at the BS. For the constraints, \eqref{powerConstraint} represents the total transmit power constraint, and \eqref{phaseConstraint} defines the phase shift constraint for each STAR-RIS element. In addition, \eqref{constraint3} and \eqref{constraint4} denote the constraints on the conservation law of energy at the STAR-RIS.
One can prove that the problem~\eqref{eq:ProbP1} is non-convex and NP-hard by exploiting the same methodology as in \cite{luo2008dynamic}. Apart from this, the coupling of transmit beamforming and STAR-RIS phase shifts makes  the problem~\eqref{eq:ProbP1}  challenging to obtain a global solution. 
\subsection{AO-based Solution of Beamforming Optimization Problem} \label{sec:AOScheme}
In this subsection, we briefly discuss how can the problem \eqref{eq:ProbP1} can be tackle by leveraging the AO-based framework. Specifically, we decompose the non-convex problem into three sub-problems corresponding to separate variables, i.e. $\mathbf{w}_k$, $\pmb{\beta}^\chi$, and $\mathbf{\Theta}^\chi$, and solve then in an iterative manner. For given $\mathbf{w}_k$ and $\pmb{\beta}^\chi$, by introducing $s_{ln}^\chi = e^{j\theta_{ln}^\chi}$, we can show that $\mathbf{h}_{kl}^H\mathrm{diag}(\pmb{\beta}_l^\chi)\mathbf{\Theta}_l^\chi\mathbf{G}_l\mathbf{w}_i=\mathbf{t}_{kil}^H\mathbf{s}_l^\chi$, where $\mathbf{t}_{kil}=(\mathrm{diag}(\mathbf{h}_{kl}^H)\mathrm{diag}(\pmb{\beta}_l^\chi)\mathbf{G}_l\mathbf{w}_i)^\ast \in \mathbb{C}^{M}$, and $\mathbf{s}_l^\chi = [s_{l1}^\chi,\cdots,s_{lM}^\chi]$. Then, the problem \eqref{eq:ProbP1} can be re-written as
\begin{subequations}\label{eq:Prob1.1}
\begin{align} 
     &\underset{\pmb{\Theta}^\chi,\pmb{\eta}}{\mathrm{maximize}} &&\sum_{k \in \mathcal{K}} \log_2(1+\eta_k) \label{eq:OptPhase}\\
    & \mbox{subject to} && \eta_k  \leq \frac{|\mathbf{t}_{kk}^H\mathbf{s}^\chi|^2}{\sum_{i\in \mathcal{K},i\neq k}|\mathbf{t}_{ki}^H\mathbf{s}^\chi|^2+\sigma^2} \forall k\in \mathcal{K},\label{eq:SlackConstraint}\\
    & && |s_{lm}^{\chi}|=1, \quad \forall l \in \mathcal{L}, m \in \mathcal{M},
\end{align}
\end{subequations}
where $\pmb{\eta} = [\eta_1,\cdots,\eta_k]^T$ stands for a slack variable, $\mathbf{s}^\chi = [\mathbf{s}^\chi_1,\cdots,\mathbf{s}^\chi_L]$, and $\mathbf{t}_{kil}=(\mathrm{diag}(\mathbf{h}_{kl}^H)\mathrm{diag}(\pmb{\beta}_l^\chi)\mathbf{G}_l\mathbf{w}_i)^\ast \in \mathbb{C}^{M}$.
To handle the non-convexity in the constraints \eqref{eq:SlackConstraint}, we introduce a slack variable $\alpha_k$ and decompose each constraint of \eqref{eq:SlackConstraint} into two following constraints
\begin{align}
    & -|\mathbf{t}_{kk}^H\mathbf{s}^\chi|^2 + \alpha_k\eta_k \leq 0  \label{eq:inequa1}\\
    & \sum_{i\in \mathcal{K},i\neq k}|\mathbf{t}_{ki}^H\mathbf{s}^\chi|^2+\sigma^2 \leq \alpha_k.\label{eq:inequa2}
\end{align}
The non-convex constraint \eqref{eq:inequa1} is relaxed by applying Taylor approximation as
\begin{align}\label{eq:inequa1Taylor}
      &-2\mathrm{Re}((\mathbf{t}_{kk}^H(\mathbf{s}^\chi)^{(n)})^H\mathbf{t}_{kk}^H(\mathbf{s}^\chi-(\mathbf{s}^\chi)^{(n)})) -|\mathbf{t}_{kk}^H(\mathbf{s}^\chi)^{(n)}|^2 \notag\\
    & + \frac{1}{4}((\alpha_k + \eta_k)^2-2(\alpha^{(n)}_k-\eta_k^{(n)})(\alpha_k-\eta_k)\notag\\
    &+(\alpha^{(n)}_k-\eta_k^{(n)})^2) \leq 0.  
\end{align} 
Next, the non-convex unit modulo constraint is relaxed by using the penalty method as 
\begin{equation}\label{eq:PenaltyObj}   
\sum_{k=1}^{K}  \log_2(1+\eta_k) + C\sum_{k=1}^{K} \sum_{l=1}^{L}\sum_{m=1}^{M}(|s_{lm}^\chi|^2-1), 
\end{equation}
where $C$ is a large positive constant. 
The non-convexity in \eqref{eq:PenaltyObj} is then relax by a convex approximation through the first order Taylor approximation as
\begin{equation}
\begin{split}
    &\sum_{k=1}^{K} \log_2(1+\eta_k) + \\
    &2C\sum_{k =1}^{K}\sum_{l=1}^{L}\sum_{m=1}^{M}\mathrm{Re}(((s_{lm}^\chi)^{(n)})^H({s}_{lm}^\chi-(s_{lm}^\chi)^{(n)})))
\end{split}
\end{equation} 
Then, the problem \eqref{eq:Prob1.1} can be recast into the following approximated convex problem:
\begin{subequations} \label{eq:SCA-P1.1}
\begin{align}
    &\underset{\mathbf{s},\eta,\alpha·}{\mathrm{maximize}} &&\sum_{k\in \mathcal{K}} \log_2(1+\eta_k) + \\
    & &&2C\sum_{k =1}^{K}\sum_{l=1}^{L}\sum_{m=1}^{M}\mathrm{Re}(((s_{lm}^\chi)^{(n)})^H({s}_{lm}^\chi-(s_{lm}^\chi)^{(n)}))) \notag\\
    &\mbox{subject to} &&|s_{lm}^\chi| \leq 1, \quad \forall l \in \mathcal{L}, m \in \mathcal{M}, \\
    & && \alpha_k \geq 0,\eqref{eq:inequa2}, \eqref{eq:inequa1Taylor}, \quad \forall k\in\mathcal{K}
\end{align}
\end{subequations}
The problem \eqref{eq:Prob1.1} can be  addressed by repeatedly solving the approximated convex problem \eqref{eq:SCA-P1.1}, employing the SCA method. The details of the SCA algorithm for solving \eqref{eq:Prob1.1} is summarized in Algorithm~\ref{alg:cap}. The convergence of Algorithm~\ref{alg:cap} is provided by the following Lemma.
\begin{lemma} \label{Lemma1}
The objective in \eqref{eq:PenaltyObj} achieved by Algorithm~\ref{alg:cap} is monotonically increasing and the variables $({\mathbf{s}}^{\chi})^{(n)}$, ${\alpha}_k^{(n)}$, ${\eta}_k^{(n)}$ converge to the points that fulfill the KKT of the problem \eqref{eq:PenaltyObj}.
\end{lemma}
\begin{IEEEproof}
    Please refer to Appendix \ref{Appx:B}
\end{IEEEproof}
Lemma \ref{Lemma1} demonstrates that by iteratively solving the approximate problem \eqref{eq:SCA-P1.1}, we can conduct a series of feasible solutions that eventually converge to the KKT solution of the problem \eqref{eq:PenaltyObj}. With a proper penalty factor $C$, we can find a feasible solution for the original problem \eqref{eq:Prob1.1} by following Algorithm~\ref{alg:cap}. It worth noting that the optimization of STAR-RIS amplitude matrix $\pmb{\beta}^{\chi}$ and beamforming vectors $\mathbf{w}_k$ can be processed in a similar topology with the optimization of $\pmb{\Theta}^\chi$. Overall, the assurance of problem convergence is underpinned by two fundamental factors. Firstly, as indicated in Lemma \ref{Lemma1}, the convergence point of the SCA
algorithm satisfies the KKT conditions of the approximated convex problem, ensuring that the objective function is non-decreasing through each iteration. Secondly, due to the power constraint and unit-modulo constraint, it is evident that the objective function is upper bound. Therefore, the convergence of the iterative optimization framework is guaranteed. In addition, the primary computational cost of the AO framework arises from solving the three sub-problems, whose complexity is in order of $\mathcal{O}(\sqrt{N}\log_2(1/\epsilon))$, where $N$ is the number of variables
and $\epsilon$ is the convergence tolerance level \cite{Miguel1998-SecondoderCone}. Overall, the total complexity of the AO algorithm is obtained as $\mathcal{O}\left(S(KLM)^{3.5}\log_2(1/(\epsilon_1\epsilon_2))+S(N_tK)^{3.5}\log_2(1/\epsilon_3)\right)$ with $S$ being the number of iterations is required to reach the KKT point of the approximated convex problem.
\begin{algorithm}[t]
\caption{SCA for solving the problem \eqref{eq:Prob1.1}} \label{alg:cap}
\begin{algorithmic}[1]
\State Initialization: $(\mathbf{s}^\chi)^{(0)}, \alpha_k^{(0)},\eta_k^{(0)}, k \in \{1,2\}$ and the iteration index $n=0$.
\State \textbf{repeat}
\State \quad Solve \eqref{eq:SCA-P1.1} to obtain $(\hat{\mathbf{s}}^\chi)^{(n)}, \hat{\alpha}_k^{(n)}, \hat{\eta}_k^{(n)}, k \in \mathcal{K}$.
\State \quad Update: \\
\quad \quad $(\mathbf{s}^\chi)^{(n+1)} = (\hat{\mathbf{s}}^\chi)^{(n)}$,$\alpha_k^{(n+1)} = \hat{\alpha}_k^{(n)}$, $\eta_k^{(n+1)} = \hat{\eta}_k^{(n)}$ \\
 \quad \quad $t := t + 1$.
\State \textbf{until} The fractional increment in the objective function in \eqref{eq:Prob1.1} remains below a specified threshold $\epsilon_1 > 0$.
\State \textbf{Return} $\mathbf{s}^{\chi *} = (\hat{\mathbf{s}}^\chi)^{(n)}, \alpha_k^* = \hat{\alpha}_k^{(n)}, \eta_k^*= \hat{\eta}_k^{(n)}, k \in \mathcal{K}$.

\end{algorithmic}
\end{algorithm}

\begin{remark}
The distributed STAR-RIS system can be addressed in a centralized manner using large matrix expressions; however, solving the resulting optimization problem incurs high computational complexity. In this work, our primary objective is to develop an efficient and scalable beamforming framework tailored for distributed STAR-RIS systems that leverage the near-far effect. To achieve this, we introduce a mathematical abstraction that models the distributed STAR-RIS system as a unified representation for beamforming optimization. This abstraction compactly expresses the distributed STAR-RIS elements and their cascaded channels, simplifying the mathematical formulation while preserving the system’s generality. Consequently, our framework enables a scalable and adaptable GNN-based solution.
\end{remark}

\section{Graph Neural Network-based Solution} \label{Sec:BGNN}
\begin{figure} [t]
    \centering
    \includegraphics[trim=0cm 0cm 0cm 0cm, clip=true, width=3.3in]{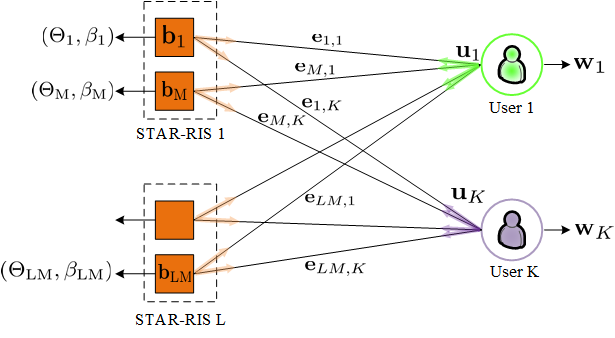}
    \caption{The graph representation of the distributed STAR-RISs systems.}
    \label{fig:BGNNArchitec}
    
\end{figure}

Due to the high computational cost, the scalability of the AO-based framework is constrained. In this section, we propose a scalable GNN-based solution to address the sum rate optimization problem under consideration.  We begin by depicting the STAR-RIS multi-user system using a heterogeneous graph. This graph comprises STAR-RIS and user vertices, effectively capturing the dynamic interplay between users and STAR-RISs within the system as illustrated in Fig.~\ref{fig:BGNNArchitec}. 

\subsection{Properties of the joint beamforming design policy}
We now show that the optimal active and passive beamforming policy enjoys permutation equivalence property. We start by re-writing the objective function in \eqref{eq:ProbP1} as 

\begin{align}\label{eq:NewSINR}
    \gamma_k &= \frac{|\sum_{l=1}^{L}\sum_{m=1}^{M}h_{klm}\phi_{lm}^{\chi}\mathbf{G}_{lm}\mathbf{w}_k|^2}{\sum_{j\in \mathcal{K},j\neq k}|\sum_{l=1}^{L}\sum_{m=1}^{M}h_{klm}\phi_{lm}^{\chi}\mathbf{G}_{lm}\mathbf{w}_j|^2+\sigma^2}\notag\\
    &= \frac{|\sum_{l=1}^{L}\sum_{m=1}^{M}\phi_{lm}^{\chi}\Tilde{\mathbf{h}}_{klm}^H\mathbf{w}_k|^2}{\sum_{j\in \mathcal{K},j\neq k}|\sum_{l=1}^{L}\sum_{m=1}^{M}\phi_{lm}^{\chi}\Tilde{\mathbf{h}}_{klm}^H\mathbf{w}_j|^2+\sigma^2},
\end{align}
where $\mathbf{h}_{kl} = [h_{kl1},\cdots,h_{klM}]^T$,  $\mathbf{G}_{lm}$ denotes the $m$-th row of $\mathbf{G}_l$,  and $\Tilde{\mathbf{h}}_{klm}^H = h_{klm}\mathbf{G}_{lm} \in \mathbb{C}^{1\times N_t}$ denoting the equivalent channel between the $m$-th element of the $l$-th STAR-RIS and the $k$-th user including the channel from BS. With the re-written  objective function, one may view the original system as a system comprising STAR-RIS and users with the corresponding propagation channels between them. We will learn a joint optimal active and passive beamforming policy as

\begin{equation} \label{eq:BFpolicy}
   \{ \mathbf{W}^\ast ,\pmb{\Phi}^{\chi \ast} \} = F(\mathbf{H}), 
\end{equation}
where $F(.)$ is the optimal beamforming policy that is needed to be learned, $\mathbf{W}^\ast = [\mathbf{w}_1^\ast,\cdots,\mathbf{w}_K^\ast]$ is the optimal beamforming matrix, $\pmb{\Phi}^{\chi \ast} = [(\pmb{\Phi}^{\chi \ast}_1)^T,\cdots,(\pmb{\Phi}^{\chi \ast}_L)^T]^T$ denotes the optimal STAR-RIS phase shift, and $\mathbf{H}$ is the equivalent channel matrix between STAR-RIS elements and users, which is expressed as
\begin{equation} \label{eq:EquivalentCSI}
    \mathbf{H}=\begin{pmatrix}  \Tilde{\mathbf{h}}_{111}&\cdots&\Tilde{\mathbf{h}}_{K11}\\
    \vdots &\vdots & \vdots\\
    \Tilde{\mathbf{h}}_{1LM}& \cdots & \Tilde{\mathbf{h}}_{KLM}
    \end{pmatrix} \in \mathbb{C}^{LM \times K}.
\end{equation} 
Before investigating the PI and PE properties of the joint active and passive beamforming optimization problem, we introduce the definition of these properties. Let us consider a multivariate function $\mathbf{Y} = f(\mathbf{X})$, and a permutation matrix $\mathbf{\Pi}$, the PI and PE properties of $f(\mathbf{X})$ is defined by the following definition.

\textit{Definition 1: For an arbitrary permutation of matrix $\mathbf{X}$, i.e, denoted by  $\mathbf{\Pi}\mathbf{X}$, if $\mathbf{Y} = f(\mathbf{\Pi}\mathbf{X})$, then $\mathbf{Y} = f(\mathbf{X})$ exhibits permutation invariance to $\mathbf{X}$. Additionally, if $\mathbf{\Pi}\mathbf{Y} = f(\mathbf{\Pi}\mathbf{X})$, then $\mathbf{Y} = f(\mathbf{X})$ is permutation equivalent to $\mathbf{X}$.}

We now investigate the properties of the beamforming optimization problem. Firstly, if the order of STAR-RISs is permuted, the optimal STAR-RIS phase shifts, and the rows of $\mathbf{H}$ are permuted to $\mathbf{\Pi}_1\pmb{\Phi}^{\chi *}$, and $\mathbf{\Pi}_1\mathbf{H}$, respectively. Moreover, when the order of users is permuted, the optimal beamforming vectors and the columns of $\mathbf{\Pi}_1\mathbf{H}$ are permuted to $\mathbf{\Pi}_2\mathbf{W}^\ast$ and $\mathbf{\Pi}_1\mathbf{H}\mathbf{\Pi}_2^T$. In contrast, as can be observed from \eqref{eq:NewSINR}, exchanging the order of STAR-RIS elements or users simply alters the cascaded paths between the BS and users and will not change the problem itself. Thus, the beamforming policy defined in \eqref{eq:BFpolicy} remains unchanged. Consequentially, we have
\begin{equation}
    \mathbf{\Pi}_1\pmb{\Phi}^{\chi *}, \mathbf{\Pi}_2\mathbf{W}^* = F(\mathbf{\Pi}_1\mathbf{H}\mathbf{\Pi}_2^T).
\end{equation}
In other words, the optimal beamforming policy is permutation equivariant. By identifying the PE  of the beamforming policy, we can design a neural network that not only simply approximates the mapping between channel information and optimal solution, but also preserves this structural property.

\subsection{Graphical Representation of STAR-RIS systems}
We now introduce a comprehensive graphical framework that sheds light on the dynamics of the considered distributed STAR-RIS multi-user system. As illustrated in Fig.~\ref{fig:BGNNArchitec},
the system may be represented as a heterogeneous graph denoted as $\mathcal{G}(\mathcal{V},\mathcal{E})$, where the STAR-RIS elements and users form two vertex sets denoted as $\mathcal{S}$ and $\mathcal{K}$, where $|\mathcal{S}| = LM$ and $|\mathcal{K}| = K$. In addition, $\mathcal{E}$ comprises the set of undirected edges $(s,k)$ connecting RIS element vertex $s$ and user vertex $k$, $\forall s \in \mathcal{S}, k \in \mathcal{K}$. To effectively capture the interaction between user vertices and RIS vertices in the wireless graph, we initially establish the characteristics for each vertex and edge.  It is suggested from \eqref{eq:NewSINR} that the equivalent channels between STAR-RISs and users are sufficient for the optimization of beamforming and phase shift vectors. To this end, we define the feature for the 
$(s,k)$ edge connecting the $s$-th RIS vertex and the $k$-th user vertex as \footnote{The channel information exploited by HGNN is the cascaded CSI between BS and users via STAR-RISs rather than the individual one. In practice, the cascaded CSI can be estimated at the BS by advanced channel estimation techniques \cite{Chen2023-RISCE}.}
\begin{equation} \label{eq:edgeFea}
        \mathbf{e}_{s,k} = [\mathrm{Re}(\mathbf{H}[s,k]), \mathrm{Im}(\mathbf{H}[s,k])] \in \mathbb{R}^{2N_t},
\end{equation}
where $\mathbf{H}[s,k]$ denotes the $(s,k)$-th element of matrix $\mathbf{H}$. 
In order to facilitate the exchange of information among vertices, we introduce a heterogeneous graph message passing (HGMP) protocol. This protocol enables the dissemination of knowledge across the entire heterogeneous
graph, allowing for the collaborative sharing of pertinent statistics required for optimizing the active and passive beamforming vectors. The HGMP inference is processed through a series of the $T$ iterations. During each iteration, every vertex communicates with its neighboring vertices, which results in an update of its internal state by processing the received messages from its adjacent vertices. Specifically, the update procedure of a vertex state during the $t$-th iteration can be defined as
\begin{itemize}
    \item \textit{User vertex update:} 
    \begin{align} 
       \mathbf{u}^{(t)}_k = \mathcal{U}^t(\mathbf{u}^{(t-1)}_k,\mathbf{w}_k^{(t-1)},\mathbf{c}_k^{(t)}) \label{eq:UserVertexComb}, 
    \end{align}
    \item \textit{RIS vertex update}
    \begin{align}
        \mathbf{b}^{(t)}_s = \mathcal{B}^t(\mathbf{b}^{(t-1)}_s,\Theta_s^{(t-1)},\beta_s^{(t-1)},\mathbf{d}_s^{(t)})\label{eq:RISVertexComb},
    \end{align}
\end{itemize}
 where $\mathbf{u}^{(t)}_k$ and $\mathbf{b}^{(t)}_s$ are the vertex feature of the $k$-th user vertex and the $s$-th RIS vertex at the $t$-th iteration, respectively. In addition, $\mathcal{U}^t(\cdot)$ and $\mathcal{B}^t(\cdot)$ are user vertex and RIS vertex combination operators. Moreover, $\mathbf{c}_i^{t}$ and $\mathbf{d}_i^{t}$ are the aggregated messages from a vertex's neighbors, which are given, respectively, by
 \begin{subequations} \label{eq:aggregation}
     \begin{align}
         \mathbf{c}_k^{(t)} = \mathrm{PL}_{s\in \mathcal{S}}\{\phi^{t}(\mathbf{b}_s^{(t-1)},\mathbf{u}_k^{(t-1)},\mathbf{e}_{s,k})\}, \label{eq:UserVertexAgg}\\
         \mathbf{d}_s^{(t)} = \mathrm{PL}_{k\in \mathcal{K}}\{\psi^{t}(\mathbf{u}_k^{(t-1)},\mathbf{b}_s^{(t-1)},\mathbf{e}_{s,k})\} \label{eq:RISVertexAgg},
     \end{align}
 \end{subequations}
 where $\phi^{t}$ and $\psi^{t}$ are the aggregation operators at the user vertices and RIS vertices, and $\mathrm{PL}(.)$ is the pooling function which is dimensional-invariant. In  \eqref{eq:UserVertexComb} and \eqref{eq:RISVertexComb}, $\mathbf{w}_k^{(t-1)}$, $\Theta_r^{(t-1)}$, and $\beta_r^{(t-1)}$ are the predicted beamforming vectors, the RIS phase shift and its amplitude at the $(t-1)$-th iteration, which are updated, respectively, as
 \begin{subequations}
     \begin{align}
         \mathbf{w}_k^{(t-1)} &= \mathcal{W}(\mathbf{u}^{(t-1)}_k), \quad \forall k\in \mathcal{K}, \label{eq:BFEstimate}\\
         \Theta_s^{(t-1)} &= \mathcal{C}(\mathbf{b}_s^{(t-1)}), \quad \forall s \in \mathcal{S}, \label{eq:PhaseEstimate}\\
         \beta_s^{(t-1)} &= \mathcal{D}(\mathbf{b}_s^{(t-1)}), \quad \forall s \in \mathcal{S}, \label{eq:AmplitudeEstimate}
     \end{align}
 \end{subequations}
 where $\mathcal{W}(\cdot)$, $\mathcal{C}(\cdot)$, and $\mathcal{D}(\cdot)$ are the mapping functions. It is worth highlighting that our proposed HGMP differs from the conventional message passing procedure  described in \cite{Guo2022-HetnetBF} and \cite{Zhang2021-PowerHetGNN}. In the conventional approach, optimized variables are only predicted at the last iteration, $T$, of the procedure. Additionally, both types of vertices utilize similar computation structures for the message generation rule. This may be sufficient in the power allocation problem considered in \cite{Guo2022-HetnetBF} where there is only one type of variable to be optimized, i.e. the power allocated at each antenna. However, in the joint beamforming design problem where various types of variables need to be jointly optimized, such a message generation rule lacks the capacity to capture the heterogeneous characteristics of the graph network. On the contrary, the proposed HGMP predicts beamforming vectors and STAR-RIS phase shifts at each iteration and integrates them into a dedicated message generation rule for each vertex type. This information will be then propagated throughout the vertices and utilized in subsequent iterations for beamforming and phase shift prediction.  The updated message passing enhances the performance of the designed graph neural network.

 The proposed HGMP hinges on the pooling function in \eqref{eq:aggregation}, enabling the dimension-invariant computation of the graph. We apply the sum operator in the paper, which is a widely recognized and efficient pooling function utilized across various applications \cite{Chowdhury2021-GNNWMMSE,Shen2021-GNNPA}. The proposed HGMP inference  is presented in Algorithm~\ref{Alg3}.

 \begin{algorithm}[t]
\caption{Proposed HGMP inference} \label{Alg3}
\begin{algorithmic}[1]
\State Initialize   $\mathbf{u}_k^{(0)}, \mathbf{b}_l^{(0)}, \forall k \in \mathcal{K}, s \in \mathcal{S}$ and $t=0$.
\For{$t \gets 1$ to $T$}  
    \begin{enumerate}
        \item \textit{User vertex update:}
        
        \quad The $k$-th user vertex aggregates its received messages from  \eqref{eq:UserVertexAgg} to generate message $\mathbf{c}_k^{(t)}$ and updates its new feature $\mathbf{u}_k^{(t)}$ from \eqref{eq:UserVertexComb} and sends it to the RIS vertex.

        \quad The $k$-th user generates its corresponding beamforming vector as in \eqref{eq:BFEstimate}.
        \item \textit{RIS vertex update:}

        \quad The $s$-th RIS vertex aggregates its received messages from  \eqref{eq:RISVertexAgg} to generate message $\mathbf{d}_s^{(t)}$ and updates its new feature $\mathbf{b}_s^{(t)}$ from \eqref{eq:RISVertexComb} and sends it to the user vertex .

        \quad The $s$-th RIS vertex generates its phase shift and amplitude as in \eqref{eq:PhaseEstimate} and \eqref{eq:AmplitudeEstimate}.
    \end{enumerate}
\EndFor
\State Return the predicted beamforming vectors, RIS phase shift and amplitude at the $T$-th iteration. 
\end{algorithmic}
\end{algorithm}
\subsection{Properties of the proposed HGMP}
This subsection discusses the key properties of the HGMP that are favorable to handle the scalable joint beamforming optimization problems.
\subsubsection{Permutation equivariance} The permutation equivariance property of the HGMP is stated in Proposition~\ref{Prop:PE}.

\begin{proposition} \label{Prop:PE}
    Let the outputs of the HGMP defined in \eqref{eq:UserVertexComb} and \eqref{eq:RISVertexComb} denote $\Phi: (\mathbf{E}) \mapsto \mathbf{U}, \mathbf{B}$, where $\mathbf{E}$ is the edge feature tensor, $\mathbf{U}$ and $\mathbf{B}$ are the outputs of the HGMP corresponding to the user and RIS vertices, i.e., vertex features at the last layer of HGMP. Also, For any $\mathbf{\Pi}_1$ and $\mathbf{\Pi}_2$  denoting  RIS and user vertex permutation matrices, respectively, we have
    \begin{equation}
       \{ \mathbf{\Pi}_1\mathbf{B}, \mathbf{\Pi}_2\mathbf{U} \} = \Phi(\mathbf{\Pi}_1 \mathbf{E} \mathbf{\Pi}_2^T), 
    \end{equation}
\end{proposition}
\begin{IEEEproof}
Refer to Appendix~\ref{Appendix:Lemma2}. 
\end{IEEEproof}

As demonstrated in the previous subsection, the active and passive beamforming policy exhibits a PE property, which indicates that an optimal solution obtained from a permuted problem corresponds to a permutation of the solution derived from the original problem. Proposition~\ref{Prop:PE} confirms that the HGMP adheres to this property. Specifically, if a GNN performs well on a particular input, it also delivers comparable performance on permutations of that input. It is important to highlight that this property is not inherently guaranteed by FCNNs or CNNs. In contrast, achieving permutation equivariance necessitates data augmentation during the training of FCNNs or CNNs, adding extra computational complexity.

\subsubsection{Scalability to different system configurations} 
Due to the dimensional constraints, both FCNNs and CNNs must have the same input and output dimension in the training and testing phases, limiting the scalability of these networks to various system configurations when the number of UEs varies. In contrast, in the HGMP, each vertex with the same type is processed by the same non-linear functions, i.e., $\mathcal{B}(\cdot), \mathbf{U}(\cdot), \phi^t(\cdot),$ $ \psi^t(\cdot)$, $\mathcal{W}(.)$, $\mathcal{C}$, and $\mathcal{D}$, whose input dimension is invariant to the number of vertices.  In addition, as shown in \eqref{eq:edgeFea} the designed feature dimension is invariant of number of users and STAR-RISs.
Thus, the HGMP can be readily adapted to different scales of the considered problem in the testing phase.
\subsection{Implementation of Heterogeneous Graph Neural Network}
\begin{figure*}
    \centering
    \begin{minipage}{0.6\textwidth}
         \centering
         \includegraphics[width=4in]{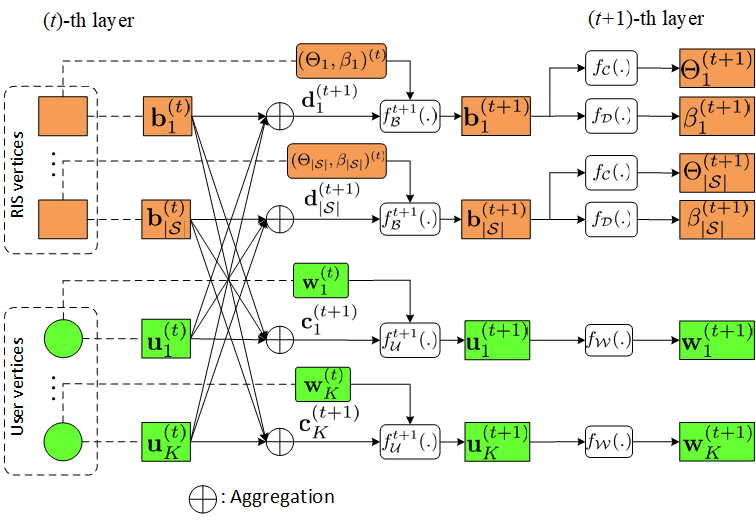}\\
         $(a)$
     \end{minipage}
     \begin{minipage}{0.3\textwidth}
		\centering
		\includegraphics[trim=0cm 0.0cm 0.0cm 0.0cm, clip=true, width=2.3in]{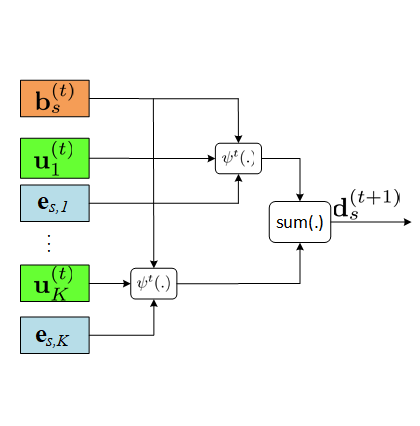} \\
		$(b)$
	\end{minipage}
    \hspace*{-0cm}
    \caption{$(a)$ The message passing procedure in the proposed HGNN between two layers. The aggregation from edge is omitted for clarity. $(b)$ The aggregation at the RIS vertex with edge feature.}
    \label{fig:HMP}
   
\end{figure*} 
In this section, we present an implementation of a beamforming heterogeneous graph neural network (BHGNN), which effectively executes the HGMP inference described in Algorithm~\ref{Alg3}.  Particularly, the BHGNN model contains $T$ layers corresponding to $T$ iterations of the proposed HGMP inference. We focus on the implementation of the vertex operators, the aggregation operators, and the mapping functions at each BHGNN layer. Instead of finding the exact structure, we adopt various FCNNs to implicitly approximate these functions. The vertex operators and aggregation operators at the $t$-th layer are designed as
\begin{align}
       \mathbf{u}^{(t)}_k &= f_{\mathcal{U}}^{t}\big(\mathbf{u}^{(t-1)}_k,\mathbf{w}_k^{(t-1)},\mathbf{c}_k^{(t)}\big), \notag\\
        \mathbf{b}^{(t)}_s &= f_{\mathcal{B}}^{t}(\mathbf{b}^{(t-1)}_s,\Theta_s^{(t-1)},\beta_s^{(t-1)},\mathbf{d}_s^{(t)}),\notag\\
        \mathbf{c}_k^{(t)} &= \sum_{s\in \mathcal{S}}(f_{\phi}^{t}(\mathbf{b}_s^{(t-1)},\mathbf{u}_k^{(t-1)},\mathbf{e}_{s,k})),\\
        \mathbf{d}_s^{(t)} &= \sum_{k\in \mathcal{K}}(f_{\psi}^{t}(\mathbf{u}_k^{(t-1)},\mathbf{b}_s^{(t-1)},\mathbf{e}_{s,k})),\notag\\
        \mathbf{w}_k^{(t)} &= f_{\mathcal{W}}(\mathbf{u}^{(t)}_k),
       \Theta_s^{(t)} = f_{\mathcal{C}}(\mathbf{b}_s^{(t)}), 
         \beta_s^{(t)} = f_{\mathcal{D}}(\mathbf{b}_s^{(t)}), \notag
\end{align}
where $f(\cdot)$ is a FCNN. The reliability of these neural networks is substantiated by the universal approximation theorem \cite{Kurt1989-Universal}, affirming that a properly constructed FCNN is able to approximate any continuous function with a sufficiently small error. Overall, the implementation of designed HGMP inference between two layers is illustrated in  Fig.~\ref{fig:HMP}. After the $T$ layers, the beamforming vectors are gathered and normalized to satisfy the transmit power constraint as follows:
\begin{align}
    \mathbf{w}_k^{(T)} &= f_{\mathcal{W}}(\mathbf{u}^{(T)}_k) \in \mathbb{R}^{2N_t}, \\
    \mathbf{W}^{(T)} &= [\mathbf{w}_1^{(T)},\cdots,\mathbf{w}_K^{(T)}] \in \mathbb{R}^{2N_t \times K},\\
    \mathbf{W}^{(T)} &= \sqrt{P_{\mathrm{max}}}\frac{\mathbf{W}^{(T)}}{||\mathbf{W}^{(T)}||_F}, \\
    \mathbf{W} &= \mathbf{W}^{(T)}(1:N_t,:) + j\mathbf{W}^{(T)}(N_t+1:2N_t,:),
\end{align}
where $\mathbf{W}(i_1:i_2,:)$ denotes a matrix constructed by taking from the $i_1$-th to the $i_2$-th row of $\mathbf{W}$.
Moreover, the STAR-RIS phase shift and amplitude in \eqref{eq:RecievedSignal} are obtained as
    \begin{align}\label{eq:PredictedGNN}
        \Theta_s^{(T)} &= f_{\mathcal{C}}(\mathbf{b}_s^{(T)}) \in \mathbb{R}^2,\quad \forall s \in \mathcal{S},\notag\\
        \mathbf{\Theta}_l^t &= e^{j2\pi[\Theta_{lM+1}^{(T)}(1),\cdots,\Theta_{(l+1)M}^{(T)}(1)]}, \quad l \in \mathcal{L}, \notag\\
        \mathbf{\Theta}_l^r &= e^{j2\pi[\Theta_{lM+1}^{(T)}(2),\cdots,\Theta_{(l+1)M}^{(T)}(2)]}, \quad l \in \mathcal{L}, \notag\\
        \beta_s^{(T)} &= f_{\mathcal{D}}(\mathbf{b}_s^{(T)}) \in \mathbb{R},\quad \forall s \in \mathcal{S}, \notag\\
        \pmb{\beta}_l^t &= \left[\sqrt{\beta_{lM+1}^{(T)}},\cdots,\sqrt{\beta_{(l+1)M}^{(T)}}\right], \quad l \in \mathcal{L}, \notag\\
        \pmb{\beta}_l^r &= \left[\sqrt{1-\beta_{lM+1}^{(T)}},\cdots,\sqrt{1-\beta_{(l+1)M}^{(T)}}\right], \quad l \in \mathcal{L}.
    \end{align}
It is worth noting that in the BHGNN architecture, all vertices employ an identical FCNN structure, which remains invariant regardless of the number of UEs or STAR-RISs. This attribute is crucial as it ensures that the BHGNN system can scale effectively, accommodating any number of users and STAR-RISs. This scalability distinguishes the BHGNN model  from traditional DL models such as CNN and FCNN,  where the system settings during the training and testing phases are required to remain unchanged.  While increasing the depth of the FCNN within BHGNN may be necessary in larger system settings, even with fixed FCNNs, any performance degradation in BHGNN would be minimal. This ability to generalize across different scenarios  will be confirmed in the simulation section.

\subsection{Training the BHGNN}
After obtaining the predicted variables as in \eqref{eq:PredictedGNN}, the  sum rate  can be readily computed according to \eqref{eq:SINR} and \eqref{eq:sumrate}. In order to train the proposed BHGNN, we define the training minimization problem on the negative expectation of the sum rate as
\begin{equation}
\underset{\mathbf{\Omega}·} 
{\mathrm{minimize}}\quad-\mathbb{E}\left[\sum_{k=1}^{K}R_k(\mathbf{W},\mathbf{\Theta}^\chi,\pmb{\beta},\mathbf{\Omega})\right], 
\end{equation}
where $\mathbf{\Omega} = \{\pmb{\omega}_{\mathcal{W}},\pmb{\omega}_{\mathcal{D}},\pmb{\omega}_{\mathcal{C}},\pmb{\omega}_{\mathcal{U}},\pmb{\omega}_{\mathcal{B}},\pmb{\omega}_{\psi},\pmb{\omega}_{\phi}\}$ is the set of the parameters of the FCNNs. The parameters update can be done using methods like the mini-batch SGD algorithm, along with its variants such as the ADAM algorithm \cite{Diederik2015-ADAM}. 
\subsection{Optimization With User QoS Constraint}
In practical multi-user systems, quality-of-service (QoS) requirements are important to ensure minimum performance levels for users. The proposed BHGNN model can be readily extended to accommodate user QoS constraints. Let us consider the per-user SINR requirement $\gamma_k \geq \bar{\Gamma}_k, \forall k \in \mathcal{K}$. By modifying the optimization problem, the Lagrangian dual framework \cite{Eisen2019-MLResourceAllo,He2022-GBLink} can be employed during the training phase to encourage the BHGNN to maximize the system sum rate while satisfying the QoS constraints for all users. Specifically, we define the modified Lagrangian loss function as
\begin{equation}
\begin{split}
     \mathcal{L}(\mathbf{W},\mathbf{\Theta}^\chi,\pmb{\beta},\mathbf{\Omega},\pmb{\lambda}) &= -\sum_{k=1}^{K}R_k(\mathbf{W},\mathbf{\Theta}^\chi,\pmb{\beta},\mathbf{\Omega}) \\
     &+ \pmb{\lambda}_k\rho_c\big(\Gamma_k-\gamma_k(\mathbf{W},\mathbf{\Theta}^\chi,\pmb{\beta},\mathbf{\Omega})\big),   
\end{split}
\end{equation}
where $\pmb{\lambda}$ is the Lagrangian multiplier associated with the QoS constraint, $\rho_c(x) \triangleq \mathrm{max}(x,0)$ is a function indicating the violation degree of the QoS constraint. Then, the BHGNN is trained to minimize the Lagrangian loss function with $\pmb{\lambda}$ adaptively updated in each epoch using the sub gradient method\cite{He2022-GBLink}. This approach ensures that the BHGNN not only maximizes system performance but also satisfies per-user QoS requirements, demonstrating its adaptability for practical applications.
\subsection{Complexity Analysis}
The computational complexity of the proposed BHGNN model consists of two key aspects: the training process and the online prediction phase. Leveraging the strong generalization capability of the BHGNN, the training process can be conducted offline and updated on a significantly longer timescale compared to the online prediction. Therefore, this work primarily focuses on the complexity of the online prediction phase, as it plays a critical role in real-time system performance. As described above, the BHGNN model comprises multiple FCNN models. For a FCNN model with $H$ hidden layers, its computational complexity is given by \cite{HaAn2021-MLCE}
\begin{equation} \label{eq:ComplexityDNN}
    C_{\mathrm{FCNN}} = \mathcal{O}\left(In_1+n_HN+\sum_{i=1}^{H-1}n_in_{i+1}\right) ,
\end{equation}
where $I$, $N$, and $n_i$ are the input size, the output size, and the number of neural in the $i$-th layers of the FCNN, respectively. Given the FCNN models designed for the HBGNN in Table \ref{DNNPara}, and by considering the number of vertices of the HBGNN, the total computational complexity of the model is on the order of $\mathcal{O}(LMN_t^2+KN_t^2+TN_t)$. Therefore, it becomes evident that the proposed HBGNN exibits lower complexity compared to the AO-based SCA algorithm.

\section{Numerical Results}\label{sec:Results}
This sections evaluates the performance of the distributed STAR-RIS-aided MU-MISO system for the proposed approaches. We also compare the distributed and centralized STAR-RIS systems under different aspects.

\subsection{Simulation Settings}
We utilize a three-dimensional (3D) Cartesian coordinates to present the positions of devices in the considered system. The BS is situated at the origin, positioned at a height of $d_H$ meter (m), where the location of the $l$-th STAR-RIS is given by $(10\cdot l,0,d_R)$. The  locations of users are uniformly distributed on the ground in the rectangular area $[20,30] \times [20,30]$ (m) in the $(x,y)$-plane for the $\mathcal{T}$ users and $[-30,-20] \times [20,30]$ (m) in the $(x,y)$-plane for the $\mathcal{R}$ users.  If not explicitly stated, we assume the number of BS antennas to be $N_t = 16$ with the transmit power budget $P_{\mathrm{max}} = 30$~dBm. The number of STAR-RISs is $L = 4$, where the number of STAR-RIS elements varies according to the scenarios. 
The channels between the BS and STAR-RISs and between STAR-RISs and users follow the Rician fading channel models as
\begin{equation} \label{eq:RiceChannel}
\begin{aligned}
     \mathbf{G}_l &= \kappa_{1,l}\left(\sqrt{\frac{\xi}{\xi + 1}}\Bar{\mathbf{G}}_l+\sqrt{\frac{1}{\xi + 1}}\Tilde{\mathbf{G}}_l\right), \\
    \mathbf{h}_{kl} &= \kappa_{2,kl}\left(\sqrt{\frac{\xi}{\xi + 1}}\Bar{\mathbf{h}}_{kl}+\sqrt{\frac{1}{\xi + 1}}\Tilde{\mathbf{h}}_{kl}\right),
\end{aligned}
\end{equation}
where $\kappa_{1,l}$ represents the distance path-loss of the channel link modeled as $32.6+36.7\log_{10}(d)$ dB, with $d$ signifies the channel link distance measured in meter, $\Tilde{\mathbf{G}}_l$ and $\Tilde{\mathbf{h}}_{kl}$ denote the non-light-of-sight (NLOS) components, which follow standard Gaussian distributions. In addition, $\Bar{\mathbf{G}}_l$ and $\Bar{\mathbf{h}}_{kl}$ denote the light-of-sight (LOS) components. We assume the BS antennas are arranged in a ULA structure, while STAR-RISs elements are arranged in the form of a UPA structure. Thus, the LoS components in \eqref{eq:RiceChannel} are expressed as the product of the UPA and ULA response vector as \cite{SZhang2020-RISCapacity}
\begin{equation} \label{eq:LoSChannel}
\begin{split}
    \Bar{\mathbf{G}}_l &= \mathbf{a}_{R_l}(\theta^A_{TR_l},\phi^A_{TR_l})\mathbf{a}_T^H(\theta^D_{TR_l}), \\
    \Bar{\mathbf{h}}_{kl} &= \mathbf{a}_{R_l}(\theta^D_{R_lk},\phi^D_{R_lk}),
\end{split}
\end{equation}
where $ \mathbf{a}_{R_l}(\theta^A_{TR_l},\phi^A_{TR_l})$ is the steering vector at the $l$-th STAR-RIS, $\mathbf{a}_T(\theta^D_{TR_l})$ is the steering vector at the BS, $(\theta^A_{TR_l}, \phi^A_{TR_l})$ denote the azimuth and elevation angle-of-arrival (AoA) at the $l$-th STAR-RIS,  
$(\theta^D_{R_lk},\phi^D_{R_lk})$ presents the azimuth and elevation angle-of-departure (AoD) from the $l$-th STAR-RIS to the $k$-th user, and $\theta^D_{TR_l}$ is the AoD from the BS to the $l$-th STAR-RIS. In \eqref{eq:LoSChannel} the steering vectors are modelled as
\begin{equation}
    \begin{split}
        &[\mathbf{a}_{R_l}(\theta^A_{TR_l},\phi^A_{TR_l})]_n = \\
        & \quad \quad e^{j\frac{2\pi d_{RIS}}{\lambda}\{i_1(n)\mathrm{sin}(\theta^A_{TR_l})\cos(\phi^A_{TR_l})+i_2(n)\sin(\phi^A_{TR_l})\}}, \\
        &[\mathbf{a}_T(\theta^D_{TR_l})]_n = e^{j\frac{2\pi d_A(n-1)\sin(\phi^A_{TR_l})}{\lambda}},
    \end{split}
\end{equation}
where $d_{RIS}$ is the distance between to consecutive STAR-RIS  elements,  $d_A = \lambda/2$ is the distance between BS antennas, with $\lambda$ denotes the carrier wavelength in meter, $i_1(n) = \mod (n-1,10),$ and $i_2(n) = \lfloor (n-1)/10 \rfloor$.  For the AO-based algorithm and the SCA method, we set the penalty factor as $C = 10^4$, the convergence tolerance levels  as $\epsilon = \epsilon_1= \epsilon_3= \epsilon_2 = 10^{-3}$. Moreover, we adopt a three-layer HBGNN model, i.e. $T=3$,  with the deployment of the FCNN presented in Table~\ref{DNNPara}. To implement the proposed GNN model, we utilize the PyTorch deep learning library \cite{pytorch}. To train the proposed neural network, 50,000 channel realizations are generated as dataset, among which 45,000 channel realizations are used for the training phase and 5,000 realizations are used for the testing phase. The neural network is trained employing the ADAM optimizer \cite{Adam-2014}, with an initial learning rate of $10^{-3}$. Throughout the training phase, the learning rate undergoes reduction every 10 epochs with a decay rate of 0.95. The training process ends when the validation loss fails to decrease consistently for five consecutive epochs.
\begin{table}[t!] 

    \centering
    \caption{Parameters of the designed fully connected layer networks}
        \begin{tabular}{*3l}
        \toprule
            Name &Shape& Activation function \\
        \midrule
            $f^1_{\phi}$, $f^1_{\psi}$ & $2+2N_t \times 2+2N_t$ & ReLU\\
            $f^2_{\phi}$, $f^3_{\phi}$,$f^2_{\psi}, f^3_{\psi}$ & $574 \times 574 $ & ReLU\\
            $f^1_{\mathcal{U}}$, $f^1_{\mathcal{B}}$& $2+2N_t \times 256 $ & ReLU\\
            $f^2_{\mathcal{U}}$, $f^2_{\mathcal{B}}$,$f^3_{\mathcal{U}}$, $f^3_{\mathcal{B}}$& $574 \times 256 $ & ReLU\\
            $f_{\mathcal{W}}$ & $256 \times 256 \times 2N_t$ & ReLU \\
            $f_{\mathcal{C}}$ & $256 \times 64 \times 2$ & Sigmoid \\
            $f_{\mathcal{D}}$ & $256 \times 64 \times 1$ & Sigmoid \\
        \bottomrule
        \end{tabular} \label{DNNPara}
     
\end{table}
To facilitate comparison, we present the performance of the following benchmarks:
\begin{enumerate}
    \item \textit{AO-SCA}: The iterative algorithm for solving problem~\eqref{eq:ProbP1} leveraging the AO-SCA framework.
    \item \textit{BHGNN}: The proposed heterogeneous graph neural network presented in Section \ref{Sec:BGNN}. Unless specified otherwise, the BHGNN is trained for the setting of $(N_t,L,M,K) = (16,4,4,8)$, and is tested with different settings to show its generalization capability robust to various settings.
    \item  
    \textit{HomoGNN}: A homogeneous GNN design that is originally proposed for a single RIS-aided system in \cite{Jiang2021-GNNRIS}. We follow the design and extend it into the distributed STAR-RIS case where the beamforming vector, STAR-RIS phase shift, and amplitude are jointly optimized.
    \item 
    \textit{CNN}: Conventional CNN design that processes the entire channel matrices $\mathbf{H}$ and $\mathbf{G}$. We follow a similar architecture in \cite{HSONG2021-RISUnsuperved} where the beamforming vector, STAR-RIS phase shift, and amplitude are jointly optimized. 
    \item HetGNN: The HetGNN model proposed in \cite{Liu2023-RISGNN} for multi-RIS
    system which is extended it to distributed STAR-RIS systems.
    \item \textit{C-STAR-RIS}: We apply the AO-SCA framework to optimize beamforming vectors, phase shift, and amplitude of the system with a single STAR-RIS located near the users. The total number of the single STAR-RIS elements is set to be  equal to the total number of that in the distributed scenario, i.e. $LM$, for a fair comparison.
    \item \textit{Random Phase shift}: We optimize beamforming vectors, while the phase shift and amplitude of the STAR-RISs are randomized.
    \item AO-SCA-EXH: We run the AO-SCA algorithm with different initialization points for a sufficiently large number and output the solution with the highest sum rate performance. Since the framework is proved to converge to the KKT point of the approximated convex problem, this scheme is expected to approximate the global optimal solution of the approximated convex problem.
\end{enumerate}

\subsection{Performance Evaluation}
We first evaluate the convergence of the BHGNN model during training. In Fig.~\ref{fig:Convergence}, we present its performance on both the training and validation datasets. Additionally, we show the performances of the AO-SCA-EXH benchmark, which serves as the performance upper bound, and the AO-SCA scheme. As observed, the BHGNN model converges after approximately 40 epochs, whereas the AO-SCA approach reaches convergence in just 10 iterations. Notably, the BHGNN model eventually surpasses the AO-SCA benchmark in validation performance. Furthermore, Fig.~\ref{fig:Convergence} illustrates that the performance of the BHGNN model closely approaches the AO-SCA-EXH benchmark, validating its efficiency in learning the optimal solution.
\begin{figure}[t]
    \centering
    \includegraphics[trim=4cm 0.5cm 7.5cm 11.5cm, clip=true, width=4.5in]{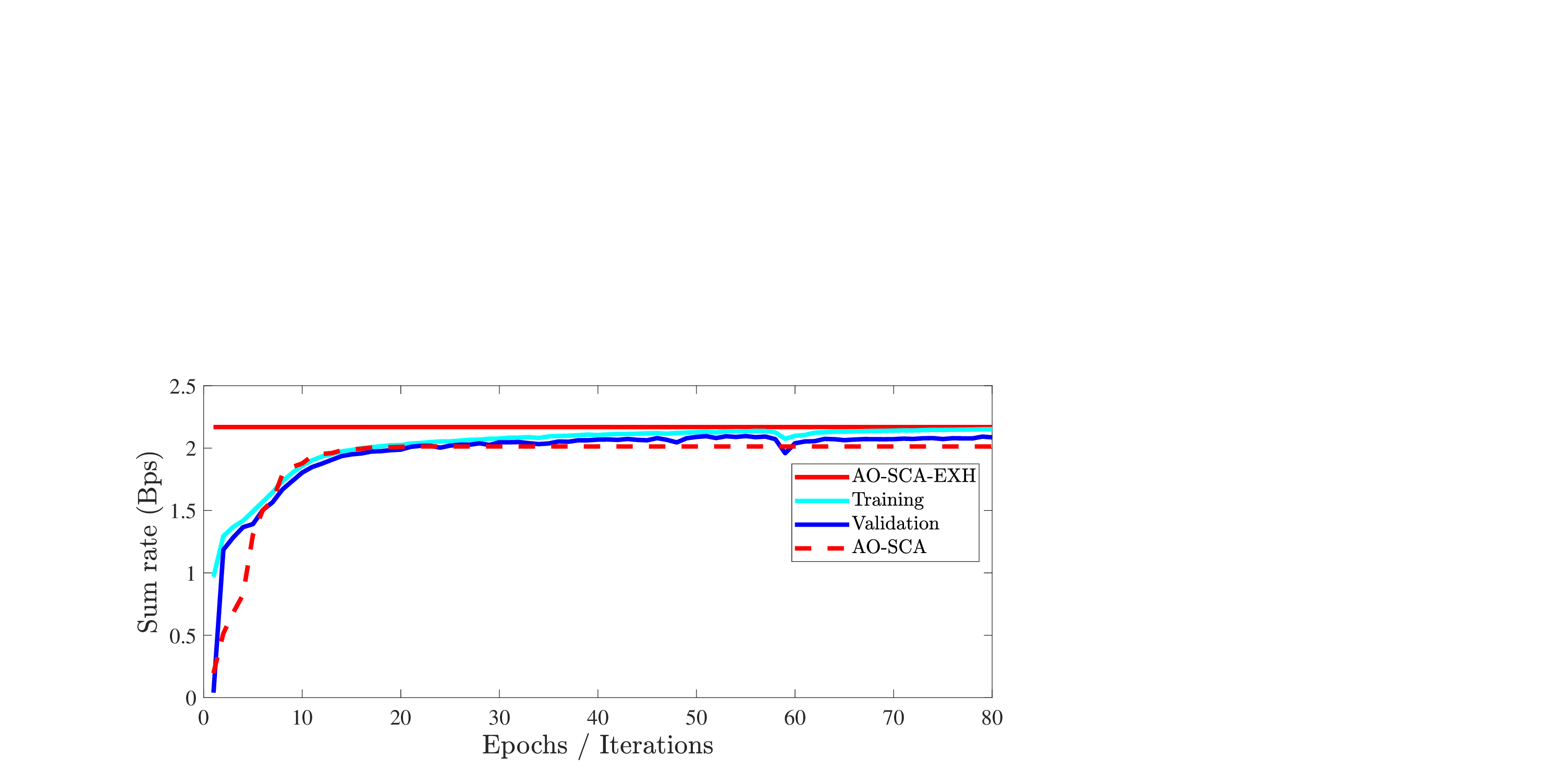}\hspace*{-3cm} 
    \caption{The training convergence of the proposed BHGNN model with $N_t = 16, K = 8, L = 4, M = 4$.}
    \label{fig:Convergence}
    \vspace{-0.5cm}
\end{figure}

Next, we compare the sum rate performance of the examined benchmarks versus the number of total STAR-RIS elements in Fig.~\ref{fig:SumratevsRIS}. To further illustrate the learning capability of the proposed BHGNN model, we also present its performance when the number of reflecting elements in the training and testing phases is identical, labeled as ``BHGNN-Robust".  As can be observed, the sum rate of all examined frameworks increases significantly with the number of STAR-RIS elements. Notably, BHGNN-Robust achieves near-optimal performance comparable to the AO-SCA-EXH scheme and outperforms the other examined benchmarks. Furthermore, as the number of STAR-RIS elements increases, the performance gap between AO-SCA and BHGNN to the AO-SCA-EXH scheme becomes larger. This is because in higher dimensions, the optimization space is extremely large.  The algorithm is, therefore, more likely to converge to a local optimal solution. In addition, while the performance of BHGNN degrades with a higher number of reflecting elements, its performance remains comparable to the AO-SCA approach, validating its strong sacalability and generalization capability.
\begin{figure}[t]
    \centering
    \includegraphics[trim=4.5cm 0.cm 7.5cm 1.5cm, clip=true, width=4.5in]{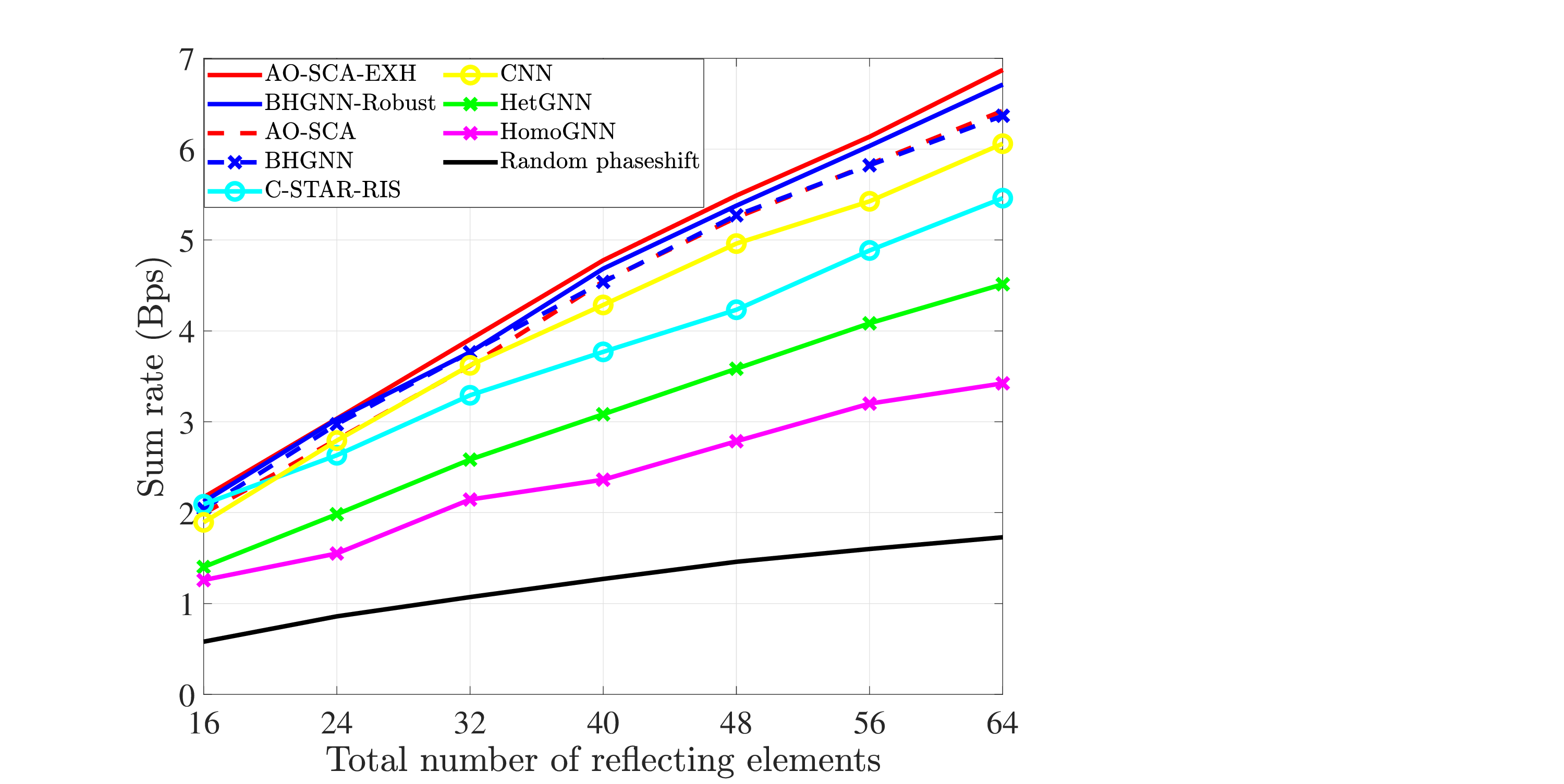}
    \caption{The sum rate performance versus total number of STAR-RIS elements of the examined methods, with $K = 8, N_t = 16$.}
    \label{fig:SumratevsRIS}
\end{figure}

\begin{table}
    \centering 
    \caption{The Generalization Performance of The BHGNN Model Via Various System Configurations}
    \begin{tabular}{ | P{1.9cm} | P{0.8cm}| P{0.8cm} | P{0.8cm}|P{0.8cm}|P{0.8cm}|} 
      \hline 
      \diagbox[width=2.3cm]{$(L,M)$}{$K$}& 8  & 10 & 12 & 14 & 16\\ 
      \hline
      $(2,4)$ & 112.7\%& 107.6\% & 104.5\% & 104.9\% & 105.1\%\\ 
      \hline
      $(2,8)$ & 112.9\% & 107.7\% & 111.2\% & 106.3\% & 108.8\% \\ 
      \hline
      $(2,12)$ & 106.1\% & 109.6\% & 106.9\% & 106.1\% & 110.6\%\\ 
      \hline
      $(3,4)$ & 106.6\%  & 107.0\%  & 101.7\%  & 100.9\%  & 100.7\% \\ 
      \hline
      $(3,8)$ & 105.1\%  & 101.1\%  & 102.4\%  & 104.7\%  & 104.0\% \\ 
      \hline
      $(3,12)$ & 102.8\%  & 103.6\%  & 102.8\%  & 104.7\%  & 103.5\%\\
      \hline
      $(4,4)$ & 107.1\%  & 108.9\%  & 103.1\%  & 102.2\%  & 104.9\% \\ 
      \hline
      $(4,8)$ & 103.1\%  & 101.7\%  & 103.8\%  & 103.8\%  & 100.6\% \\ 
      \hline
      $(4,12)$ & 98.7\%  & 99.5\%  & 99.9\%  & 100.1\%  & 101.2\%\\
      \hline
    \end{tabular} \label{Tab:BHGNNGeneralization}
    \vspace{-0.5cm}
\end{table}
 Furthermore, as seen in Fig.~\ref{fig:SumratevsRIS}, the CNN, HomoGNN, and HetGNN models underperform compared to BHGNN, underscoring the superior effectiveness of BHGNN over existing DL benchmarks. Notably, the CNN benchmark achieves better performance than both the HomoGNN and HetGNN models. This is because neither HomoGNN nor HetGNN captures the correct structural properties of the optimal beamforming policy, limiting their ability to learn the optimal solution. Specifically, the HomoGNN model generates RIS vertex features while ignoring the unique connections between STAR-RISs and users. As a result, HomoGNN fails to maintain the PE property in relation to STAR-RISs. Additionally, as shown in the Appendix \ref{Sec:ApdxC}, the predicted beamforming matrix by HetGNN  does not preserve PE, resulting in a potential performance degradation \cite{Guo2022-HetnetBF}. Furthermore, it is important to note that since the CNN, HomoGNN, and HetGNN designs are constrained by the system configuration, they must be re-trained whenever the number of RIS elements changes. 
 
To further validate the generalization capability of the proposed BHGNN model, Table~\ref{Tab:BHGNNGeneralization} presents its relative sum rate performance normalized to that of the AO-SCA scheme across various system configurations, including the number of users, STAR-RISs, and reflecting elements. As shown, the proposed BHGNN demonstrates strong generalization capability, maintaining robust performance under different system settings and achieving results comparable to the AO-based algorithm. This generalization ability significantly enhances the practical applicability of the proposed BHGNN in real-world environments.
\begin{figure}[t]
    \centering
    \includegraphics[trim=4.0cm 1cm 7.0cm 4cm, clip=true, width=4.5in]{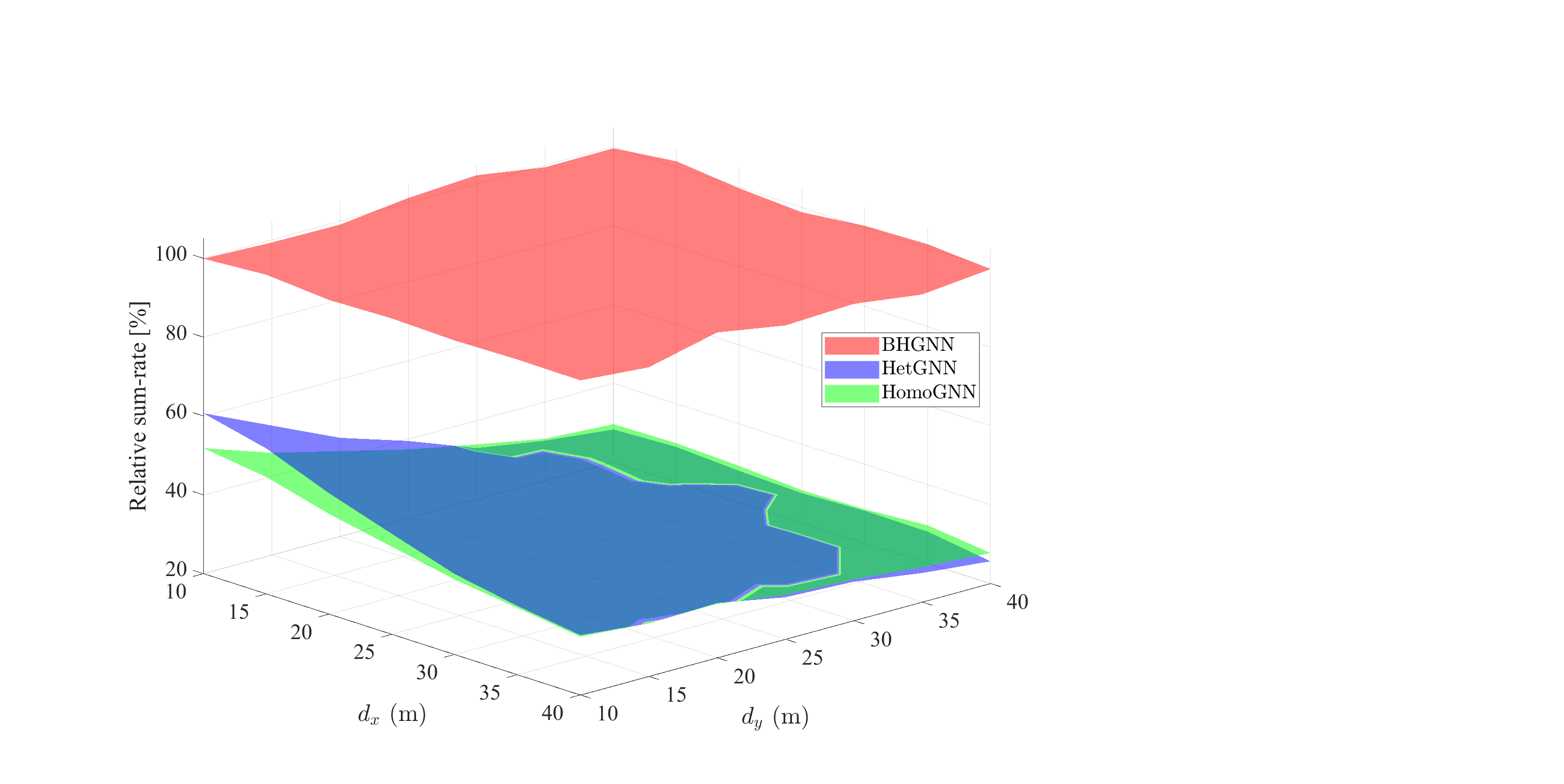}
    \caption{ The sum rate performance versus different user location's density, with $N_t = 16$, $L = 4$, $M = 10$, $K = 8$.}
    \label{fig:SumratevsUerlocation}
    \vspace{-0.3cm}
\end{figure}
\begin{figure}[t]
    \centering
    \includegraphics[trim=3cm 0.cm 7.5cm 1.5cm, clip=true, width=4.5in]{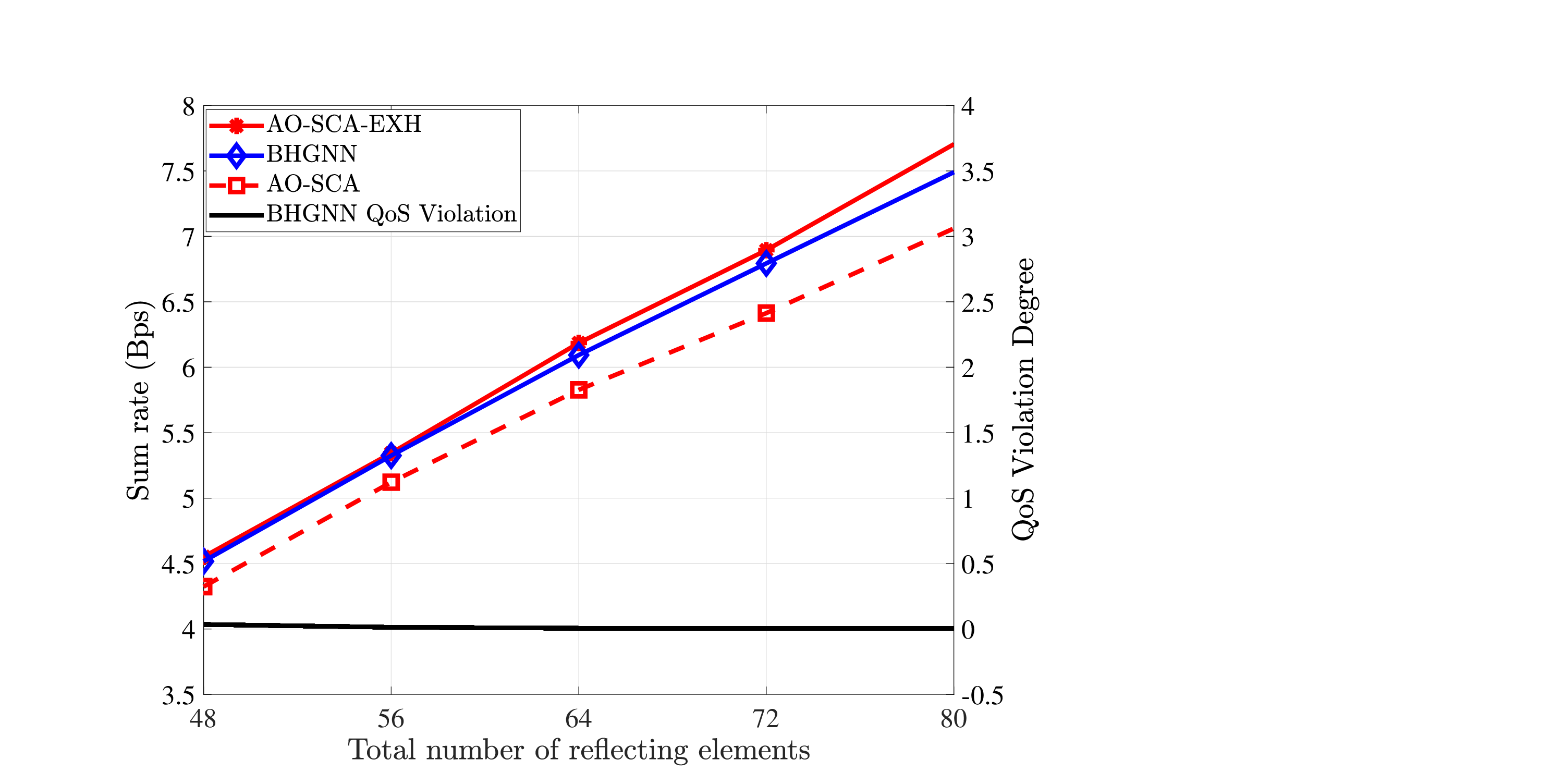}
    \caption{The performance of BHGNN model versus total number of STAR-RIS elements under QoS constraints, with $N_t = 16,K = 4, \bar{\Gamma}_k = 1$ dB.}
    \label{fig:SumratevsRISQoS}
    \vspace{-0.3cm}
\end{figure}  
Fig.~\ref{fig:SumratevsUerlocation} illustrates the relative sum rate performance obtained by the proposed BHGNN and benchmark models as a function of different user location densities. The relative sum rate is defined as the sum rate normalized by that of the AO-SCA algorithm. Specifically, the DL models are first trained as users are uniformly distributed in the rectangular area of $(x,y) = [\pm 20, \pm 30] \times [20,30] $ (m). We then vary the area of the rectangle in the testing phase as $(x,y) = [20, 20 + d_x] \times [20,20+d_y]$ (m) for $\mathcal{T}$ users and $(x,y) = [-20 - d_x, -20] \times [20,20 + d_y]$ (m) for $\mathcal{R}$ users, where $d_x$ and $d_y$ denote the range of the location rectangle. Although most of the user locations are unseen in the testing phase, the designed BHGNN obtains almost the same sum rate compared to the AO-SCA algorithm. In contrast, HomoGNN and HetGNN yield significant performance losses. Interestingly, HetGNN has a sharper performance degradation compared to HomoGNN, especially in larger network areas. The reason is that HetGNN fails to preserve the PE of optimal beamforming policy with respect to user's order.  Overall, the results validate the scalability to different network sizes of our proposed BHGNN model which can learn the universal optimization rule which allows it to generalize well robust to different system configurations.
  \begin{table}
	    \centering
	    \caption{Average Running Time of the AO-SCA and BHGNN Approaches}
    	\begin{tabular}{ | P{2.5cm} | P{1.6cm}| P{1.6cm} | P{1.5cm}|} 
          \hline
          Parameters& AO-SCA  & BHGNN \\ 
           
          \hline
          $LM = 16, K = 4$ & 177.4 ms & 25.7 ms\\ 
          \hline
          $LM = 16, K = 8$ & 288.1 ms & 29.3 ms\\ 
          \hline
          $LM = 16, K = 16$ & 821.5ms & 32.4 ms\\ 
          \hline
          $LM = 64, K = 4$ & 203.1 ms & 26.6 ms\\ 
          \hline
          $LM = 64, K = 8$ & 414.1 ms & 39.5 ms\\ 
          \hline
          $LM = 64, K = 16$ & 1271.7 ms & 43.3 ms\\ 
          \hline
        \end{tabular} \label{Running-Time}
        \vspace{-0.5cm}
    \end{table}

In Fig.~\ref{fig:SumratevsRISQoS}, we present the performance of the BHGNN model under per-user SINR QoS constraints versus the number of total STAR-RIS elements. In addition, the QoS constraint violation degree of BHGNN is also illustrated to validate its feasibility.   Fig.~\ref{fig:SumratevsRISQoS} demonstrates that while the BHGNN shows negligible degree of SINR constraint violations, its performance is near optimal  compared to the AO-SCA-EXH, validating the strong capability of the BHGNN.

Finally, we compare the computational cost of the two proposed schemes in terms of average CPU running time in Table~\ref{Running-Time}. As illustrated, the BHGNN exhibits significantly slower running time compared to the AO-SCA. Additionally, as the total number of STAR-RIS coefficients and users increases, the complexity of the BHGNN  slightly increases, while the running time of AO-SCA grows drastically. This result aligns with the complexity analysis presented in the previous sections. 
\vspace{-0.3cm}
\section{Conclusion}\label{sec:Conclusion}
In this paper, we delved into the joint optimization of active and passive beamforming in a distributed STAR-RIS assisted MU-MISO communication networks to maximize the overall sum rate. We first introduced a AO-based SCA algorithm to tackle the non-convex problem, which guaranteed to converge to the KKT solution. To enable to scalability, we further proposed a novel HGNN model to directly optimize the beamforming vectors and STAR-RIS elements, namely BHGNN. Particularly, we modeled each user and STAR-RIS coefficient as vertices, while the effective channel information was exploited as edges connecting them. Numerical results showed that the proposed BHGNN model can achieve a comparable sum rate performance to AO-based benchmark, and it can generalize well across various system configuration, e.g. different number of users, STAR-RISs as well as STAR-RIS coefficients. Furthermore, the BHGNN model requires significantly lower computational complexity compared to the AO-based benchmark, rendering it feasible for practical system implementations. 
\appendices
\section{Proof of Lemma \ref{Lemma1}} \label{Appx:B}
We first show that Algorithm 1 is guaranteed to converge by following standard arguments for the general framework of SCA \cite{Scutari2017-SCA}. Specifically, the following properties of Algorithm \ref{alg:cap} are satisfied
\begin{itemize}
    \item A feasible solution to \eqref{eq:SCA-P1.1} at the $n$-th iteration is also feasible to \eqref{eq:PenaltyObj} and \eqref{eq:SCA-P1.1} at the $(n+1)$-th iteration.
    \item The objective function generated by Algorithm \ref{alg:cap} is monotonically increasing.
\end{itemize}
Let us denote the objective function of \eqref{eq:PenaltyObj} as $f(\mathbf{s},\pmb{\eta})$, the objective function of \eqref{eq:SCA-P1.1} at the $n$-th iteration as $\hat{f}(\mathbf{s},\pmb{\eta};\mathbf{s}^{(n)})$, the LHS of  \eqref{eq:inequa1} as $g(\mathbf{s},\eta_k,\alpha_k)$, and $\hat{g}(\mathbf{s},\eta_k,\alpha_k;\mathbf{s}^{(n)},\eta_k^{(n)},\alpha_k^{(n)})$ is the LHS of \eqref{eq:inequa1Taylor}. Suppose $(\mathbf{s},\eta_k,\alpha_k)$ is feasible to \eqref{eq:SCA-P1.1}, i.e. $\hat{g}(\mathbf{s},\eta_k,\alpha_k;\mathbf{s}^{(n)},\eta_k^{(n)},\alpha_k^{(n)})\leq 0$, it holds that $g(\mathbf{s},\eta_k,\alpha_k) \stackrel{(a)}{\leq }\hat{g}(\mathbf{s},\eta_k,\alpha_k;\mathbf{s}^{(n)},\eta_k^{(n)},\alpha_k^{(n)}) \leq 0$. Inequality $(a)$ holds because $\hat{g}(\mathbf{s},\eta_k,\alpha_k;\mathbf{s}^{(n)},\eta_k^{(n)},\alpha_k^{(n)})$ is approximation of concave terms in $g(\mathbf{s},\eta_k,\alpha_k)$. Thus, $(\mathbf{s},\eta_k,\alpha_k)$ is also feasible to \eqref{eq:PenaltyObj}. In addition, since $(\hat{\mathbf{s}}^{(n)},\hat{\eta}_k^{(n)},\hat{\alpha}_k^{(n)})$ is an optimal solution to \eqref{eq:SCA-P1.1} at the $n$-th iteration, we have $\hat{g}(\hat{\mathbf{s}}^{(n)},\hat{\eta}_k^{(n)},\hat{\alpha}_k^{(n)};\mathbf{s}^{(n)},\eta_k^{(n)},\alpha_k^{(n)}) \leq 0$. Further, since $g(\hat{\mathbf{s}}^{(n)},\hat{\eta}_k^{(n)},\hat{\alpha}_k^{(n)}) \leq \hat{g}(\hat{\mathbf{s}}^{(n)},\hat{\eta}_k^{(n)},\hat{\alpha}_k^{(n)};\mathbf{s}^{(n)},\eta_k^{(n)},\alpha_k^{(n)})$, it holds that $\hat{g}(\hat{\mathbf{s}}^{(n)},\hat{\eta}_k^{(n)},\hat{\alpha}_k^{(n)};\hat{\mathbf{s}}^{(n)},\hat{\eta}_k^{(n)},\hat{\alpha}_k^{(n)})=g(\hat{\mathbf{s}}^{(n)},\hat{\eta}_k^{(n)},\hat{\alpha}_k^{(n)})\leq 0$, which implies $(\hat{\mathbf{s}}^{(n)},\hat{\eta}_k^{(n)},\hat{\alpha}_k^{(n)})$ is also feasible to \eqref{eq:SCA-P1.1} at the $(n+1)$-th iteration. Now we show that objectives generated by Algorithm \ref{alg:cap} is monotonically increasing. Since $f(\mathbf{s},\pmb{\eta})$ is convex in $\mathbf{s}$, we have $f(\hat{\mathbf{s}}^{(n)},\hat{\pmb{\eta}}^{(n)}) \geq \hat{f}(\hat{\mathbf{s}}^{(n)},\hat{\pmb{\eta}}^{(n)};\hat{\mathbf{s}}^{(n-1)}) \geq \hat{f}(\hat{\mathbf{s}}^{(n-1)},\hat{\pmb{\eta}}^{(n-1)};\hat{\mathbf{s}}^{(n-1)}) = f(\hat{\mathbf{s}}^{(n-1)},\hat{\pmb{\eta}}^{(n-1)})$. Furthermore, due to the power modulo constraints, the objective function of \eqref{eq:PenaltyObj} is bounded from above. Therefore, Algorithm \ref{alg:cap} is guaranteed to converge. Now let $(\mathbf{s}^*,\pmb{\eta}^*,\pmb{\alpha}^*)$ denote the limit point generated by Algorithm \eqref{alg:cap}. Since it is the optimal solution for the convex problem \eqref{eq:SCA-P1.1}, it is a KKT point to problem \eqref{eq:SCA-P1.1}. Now by using the fact that $f(\mathbf{s}^*,\pmb{\eta}^*) = \hat{f}(\mathbf{s}^*,\pmb{\eta}^*;\mathbf{s}^*,\pmb{\eta}^*)$, $g(\mathbf{s}^*,\pmb{\eta}^*,\pmb{\alpha}^*) = \hat{g}(\mathbf{s}^*,\pmb{\eta}^*,\pmb{\alpha}^*;\mathbf{s}^*,\pmb{\eta}^*,\pmb{\alpha}^*)$, and $\nabla g(\mathbf{s}^*,\pmb{\eta}^*,\pmb{\alpha}^*) = \nabla \hat{g}(\mathbf{s}^*,\pmb{\eta}^*,\pmb{\alpha}^*;\mathbf{s}^*,\pmb{\eta}^*,\pmb{\alpha}^*)$, we can easily show that $(\mathbf{s}^*,\pmb{\eta}^*,\pmb{\alpha}^*)$ is also a KKT point for \eqref{eq:PenaltyObj}.

\section{Proof of Proposition \ref{Prop:PE}}
\label{Appendix:Lemma2}
We represent the input features of user vertex $k$ and RIS vertex $s$ in the original graph as $\mathbf{u}_k^{(0)}$ and $\mathbf{b}_r^{(0)}$, respectively, the edge feature connecting RIS vertex $s$ and user vertex $k$ as $\mathbf{e}_{s,k}$, and the outputs of the $t$-th layer as $\mathbf{u}_k^{(t)}$ and $\mathbf{b}_r^{(t)}$. These variables in the permuted graph are correspondingly denoted as $\hat{\mathbf{u}}_k^{(0)}$, $\hat{\mathbf{b}}_r^{(0)}$, $\hat{\mathbf{e}}_{s,k}$, $\hat{\mathbf{u}}_k^{(t)}$, and $\hat{\mathbf{b}}_r^{(t)}$. At the initial stage of the graph, we have
 \begin{align} \label{eq:intialPermution}
         \hat{\mathbf{e}}_{\pi_1(s),\pi_2(k)} &= \mathbf{e}_{s,k}, \quad \hat{\mathbf{b}}_{\pi_1(s)}^{(0)} = \mathbf{b}_{s}^{(0)},  \hat{\mathbf{u}}_{\pi_2(k)}^{(0)} = \mathbf{u}_{k}^{(0)}, \\
         \mathcal{S}(\pi_1(s)) &= \{\pi_1(s), s \in \mathcal{S}\},  \mathcal{K}(\pi_2(k)) = \{\pi_2(k), k \in \mathcal{K}\},\notag
 \end{align}
 where $\pi_1$ and $\pi_2$ are the permutation operator on the RIS vertices and user vertices. Specifically, these operators are defined as 
 \begin{equation}
     \begin{split}
         \pi_1 : [LM] \rightarrow [LM] \quad &\text{with} \quad [LM] = \{1,\ldots LM\} \\
         \pi_2 : [K] \rightarrow [K] \quad &\text{with} \quad [K] = \{1,\ldots, K\}
     \end{split}
 \end{equation}
 For given permutation operators $\pi_1$, $\pi_2$ and $n$, we prove that $\mathbf{u}^{(n)}_k = \hat{\mathbf{u}}^{(n)}_{\pi_2(k)}, \forall k$, and $\mathbf{b}^{(n)}_s = \hat{\mathbf{b}}^{(n)}_{\pi_1(s)}, \forall n$ by induction. In the case of $n = 0$, the proof directly follows  \eqref{eq:intialPermution}. We now assume that the result holds with $ n = 1,\ldots,S-1$, that is
 \begin{align} \label{eq:InductionAssump}
        \mathbf{u}^{(n)}_k &= \hat{\mathbf{u}}^{(n)}_{\pi_2(k)}, \quad \mathbf{b}^{(n)}_s = \hat{\mathbf{b}}^{(n)}_{\pi_1(s)}, \hat{\mathbf{e}}_{\pi_1(s),\pi_2(k)}=\mathbf{e}_{s,k}, \notag\\
        \mathcal{S}(\pi_1(s)) &= \{\pi_1(s), s \in \mathcal{S}\}, \mathcal{K}(\pi_2(k)) = \{\pi_2(k), k \in \mathcal{K}\},\notag\\
        \forall n &= 1,\ldots,S-1.
 \end{align}
 We prove that the results hold with $n = S$. Following the HGMP update rule, the outputs of vertices at the $S$-th layer are 
 \begin{align}\label{eq:inductionProve}
      &\mathbf{u}^{(S)}_k = \mathcal{U}^S\left(\mathbf{u}^{(S-1)}_k,\mathcal{W}(\mathbf{u}^{(S-1)}_k),\right.\notag\\
      &\left.\mathrm{PL}_{s\in \mathcal{S}}\left\{\phi^{S}(\mathbf{b}_s^{(S-1)},\mathbf{u}_k^{(S-1)},\mathbf{e}_{s,k})\right\}\right), \notag\\
      &\hat{\mathbf{u}}^s_{\pi_2(k)} = \mathcal{U}^S\left(\mathbf{u}^{(S-1)}_{\pi_2(k)},\mathcal{W}(\mathbf{u}^{(S-1)}_{\pi_2(k)}),\right.\notag\\
      &\quad \quad \left.\mathrm{PL}_{s\in \mathcal{S}(\pi_1(s))}\left\{\phi^{S}(\mathbf{b}_r^{(S-1)},\mathbf{u}_{\pi_2(k)}^{(S-1)},\mathbf{e}_{s,\pi_2(k)})\right\}\right),\notag\\
      &\mathbf{b}^{(S)}_s = \mathcal{B}^S\left(\mathbf{b}^{(S-1)}_s,\mathcal{C}(\mathbf{b}_s^{(S-1)}),\mathcal{D}(\mathbf{b}_s^{(S-1)}),\right.\notag\\
      &\quad \quad \left.\mathrm{PL}_{k\in \mathcal{K}}\left\{\psi^{S}(\mathbf{u}_k^{(S-1)},\mathbf{b}_s^{(S-1)},\mathbf{e}_{s,k})\right\}\right)\notag\\
      &\hat{\mathbf{b}}_{\pi_1(s)}^{(S)} = \mathcal{B}^S\left(\mathbf{b}^{(S-1)}_{\pi_1(s)},\mathcal{C}(\mathbf{b}_{\pi_1(s)}^{(S-1)}),\mathcal{D}(\mathbf{b}_{\pi_1(s)}^{(S-1)}),\right.\notag\\
      &\quad \quad \left.\mathrm{PL}_{k\in \mathcal{K}(\pi_2(k))}\left\{\psi^{S}(\mathbf{u}_k^{(S-1)},\mathbf{b}_{\pi_1(s)}^{(S-1)},\mathbf{e}_{{\pi_1(s)},k})\right\}\right).
 \end{align}
Plugging \eqref{eq:InductionAssump} into \eqref{eq:inductionProve}, we have $\mathbf{u}_k^{(S)} = \hat{\mathbf{u}}_{\pi(k)}^{(S)}$ and $\mathbf{b}_s^{(S)} = \hat{\mathbf{b}}_{\pi(s)}^{(S)}$. We recall that for the original graph, the output at RIS vertices is $\mathbf{U} = [\mathbf{u}_1,\ldots,\mathbf{u}_{|\mathcal{S}|}]^T$, while that of the permuted graph is $\hat{\mathbf{U}} = [\hat{\mathbf{u}}_1,\ldots,\hat{\mathbf{u}}_{|\mathcal{S}|}]^T$. Therefore, we have $\mathbf{U} = \pi_1 \star \hat{\mathbf{U}}$, where $\pi_1 \star \mathbf{A}$ is defined as $(\pi_1 \star \mathbf{A})_{(\pi_1(i))} = \mathbf{A}_{(i)}$ for any matrix $\mathbf{A}$. Similarly, we have $\mathbf{B} = \pi_2 \star \hat{\mathbf{B}}$, where  $(\pi_2 \star \hat{\mathbf{B}})_{\pi_2(i)} = \hat{\mathbf{B}}_{(s)}$. Since permutation of the output of the original graph is the output of the permuted graph, the proposition~\ref{Prop:PE} is proved.

\section{Property of beamforming policy in \cite{Liu2023-RISGNN}} \label{Sec:ApdxC}
 
We recall the message passing in \cite{Liu2023-RISGNN}, in which the vertex's feature at the $t$-th layer is updated as
\begin{equation}
\begin{split}
    \mathbf{v}_{u_k}^{(t)} &= \rho_{u_k}(\pmb{\xi}_{ru_k}^{(t)},\pmb{\xi}_{uu_k}^{(t)})+ \mathbf{v}_{u_k}^{(t-1)},  \\
    \mathbf{v}_{r_i}^{(t)} &= \rho_{r_i}(\pmb{\xi}_{rr_i}^{(t)},\pmb{\xi}_{ur_i}^{(t)}) + \mathbf{v}_{r_i}^{(t-1)},
\end{split}
\end{equation}
where $\mathbf{v}_{u_k}^{(t)}$ and $\mathbf{v}_{r_i}^{(t)}$ represent the user and the RIS vertex features, respectively, and
\begin{align}
        \pmb{\xi}_{ru_k}^{(t)} &= \phi_{1}(\rho_{ru_k}(\pmb{\xi}_{r_1}^{(t)},\cdots,\pmb{\xi}_{r_L}^{(t)}),\mathbf{v}_{u_k}^{(t-1)}),\\
        \pmb{\xi}_{uu_k}^{(t)} &= \rho_{u_k}(\phi_{2}(\pmb{\xi}_u^{(t)}),\phi_{3}(\pmb{\xi}_{u_k}^{(t)},\mathbf{v}_{u_k}^{(t-1)}))\\
        \pmb{\xi}_{u}^{(t)} &= \rho_{u_k}(\pmb{\xi}_{u_1}^{(t)},\cdots,\pmb{\xi}_{u_K}^{(t)}),\\      
        \pmb{\xi}_{rr_i}^{(t)} &= \phi_4(\pmb{\xi}_{r_i}^{(t)},\mathbf{v}_{r_i}^{(t-1)}),\\
        \pmb{\xi}_{ur_i}^{(t)} &= \phi_5(\rho_{r_ri}(\pmb{\xi}_{u_1}^{(t)},\cdots,\pmb{\xi}_{u_K}^{(t)}),\mathbf{v}_{r_i}^{(t-1)}),\\
        \pmb{\xi}_{u_k}^{(t)} &= \phi_{6}(\mathbf{v}_{u_k}^{(t-1)}), \pmb{\xi}_{r_i}^{(t)} = \phi_{7}(\mathbf{v}_{r_i}^{(t-1)}),
\end{align}
where $\rho(.)$ denotes the mean pooling function, $\phi_i$ is a trainable FCNN model. Following the similar topology in Appendix \ref{Appendix:Lemma2}, we can show that
\begin{equation} \label{eq:PEofBenchmark}
    \mathbf{v}_{u_k}^{(t)} = \hat{\mathbf{v}}_{\pi_1(u_k)}^{(t)}, \quad \mathbf{v}_{r_i}^{(t)} = \hat{\mathbf{v}}_{\pi_2(r_i)}^{(t)},\forall t = 1,\cdots,T.
\end{equation}
We now investigate the property of the designed beamforming in \cite{Liu2023-RISGNN}. Specifically, the beamforming matrix is obtained at the last layer of GNN as
\begin{equation}
    \begin{split}
        \mathbf{W} &= \phi_b(\mathbf{v}_b^{(T)}), \mathbf{v}_b^{(T)} = \rho_b(\pmb{\xi}_{ub}^{(T)},\pmb{\xi}_{rb}^{(T)}),\\
        \pmb{\xi}_{ub}^{(T)} &= \phi_{ub}(\mathbf{v}_{u_1}^{(T)},\cdots,\mathbf{v}_{u_K}^{(T)}),\\
        \pmb{\xi}_{rb}^{(T)} &= \phi_{rb}(\mathbf{v}_{r_1}^{(T)},\cdots,\mathbf{v}_{r_L}^{(T)}),
    \end{split}
\end{equation}
where $\phi_b(.)$ is a FCNN. Let $\hat{\mathbf{W}}$ and $\hat{\mathbf{v}}_b^{(T)}$ denote the beamforming matrix and output of the BS vertex in the permuted graph, respectively. We have
\begin{align}
    &\hat{\mathbf{W}} = \phi_b(\hat{\mathbf{v}}_b^{(T)}) = \phi_b\bigg(\rho_b\Big(\phi_{ub}\big(\hat{\mathbf{v}}_{\pi_1(u_1)}^{(T)},\cdots,\hat{\mathbf{v}}_{\pi_1(u_K)}^{(T)}),\notag\\
    &\quad \phi_{rb}(\hat{\mathbf{v}}_{\pi_2(r_1)}^{(T)},\cdots,\hat{\mathbf{v}}_{\pi_2(r_L)}^{(T)}\big)\Big)\bigg)\\
    & \stackrel{(a)}{=}\phi_b\bigg(\rho_b\Big(\phi_{ub}\big(\mathbf{v}_{u_1}^{(T)},\cdots,\mathbf{v}_{u_K}^{(T)}) \phi_{rb}(\mathbf{v}_{r_1}^{(T)},\cdots,\mathbf{v}_{r_L}^{(T)}\big)\Big)\bigg) = \mathbf{W},\notag
\end{align}
where $(a)$ is obtained by using \eqref{eq:PEofBenchmark}. That is, we have $\mathbf{W}= f(\pi_1\star \mathbf{X})$, where $\mathbf{X}$ is the input of the HetGNN. Following Definition 1, it can be observed that the beamforming matrix obtained in \cite{Liu2023-RISGNN} fails to preserve PE property.

\bibliographystyle{IEEEtran}
\bibliography{refs}

\begin{thebibliography}{10}
\providecommand{\url}[1]{#1}
\csname url@samestyle\endcsname
\providecommand{\newblock}{\relax}
\providecommand{\bibinfo}[2]{#2}
\providecommand{\BIBentrySTDinterwordspacing}{\spaceskip=0pt\relax}
\providecommand{\BIBentryALTinterwordstretchfactor}{4}
\providecommand{\BIBentryALTinterwordspacing}{\spaceskip=\fontdimen2\font plus
\BIBentryALTinterwordstretchfactor\fontdimen3\font minus
  \fontdimen4\font\relax}
\providecommand{\BIBforeignlanguage}[2]{{%
\expandafter\ifx\csname l@#1\endcsname\relax
\typeout{** WARNING: IEEEtran.bst: No hyphenation pattern has been}%
\typeout{** loaded for the language `#1'. Using the pattern for}%
\typeout{** the default language instead.}%
\else
\language=\csname l@#1\endcsname
\fi
#2}}
\providecommand{\BIBdecl}{\relax}
\BIBdecl

\bibitem{WSaad2019-6GVision}
W.~Saad, M.~Bennis, and M.~Chen, ``A vision of {6G} wireless systems:
  Applications, trends, technologies, and open research problems,'' \emph{IEEE
  Network}, vol.~34, no.~3, pp. 134--142, 2020.

\bibitem{Ebasar2019-RISintro}
E.~Basar, M.~Di~Renzo, J.~De~Rosny, M.~Debbah, M.-S. Alouini, and R.~Zhang,
  ``Wireless communications through reconfigurable intelligent surfaces,''
  \emph{IEEE Access}, vol.~7, pp. 116\,753--116\,773, 2019.

\bibitem{Mu2022-STARRIS}
X.~Mu, Y.~Liu, L.~Guo, J.~Lin, and R.~Schober, ``Simultaneously transmitting
  and reflecting {(STAR)} {RIS} aided wireless communications,'' \emph{IEEE
  Transactions on Wireless Communications}, vol.~21, no.~5, pp. 3083--3098,
  2022.

\bibitem{Wang2023-STARRISNOMA}
T.~Wang, F.~Fang, and Z.~Ding, ``Joint phase shift and beamforming design in a
  multi-user {MISO} {STAR-RIS} assisted downlink {NOMA} network,'' \emph{IEEE
  Transactions on Vehicular Technology}, vol.~72, no.~7, pp. 9031--9043, 2023.

\bibitem{Wu2021-STARRISCoverage}
C.~Wu, Y.~Liu, X.~Mu, X.~Gu, and O.~A. Dobre, ``Coverage characterization of
  {STAR-RIS} networks: {NOMA} and {OMA},'' \emph{IEEE Communications Letters},
  vol.~25, no.~9, pp. 3036--3040, 2021.

\bibitem{Papaza2021-MulRISCoverage}
A.~Papazafeiropoulos, C.~Pan, A.~Elbir, P.~Kourtessis, S.~Chatzinotas, and
  J.~M. Senior, ``Coverage probability of distributed {IRS} systems under
  spatially correlated channels,'' \emph{IEEE Wireless Communications Letters},
  vol.~10, no.~8, pp. 1722--1726, 2021.

\bibitem{Wang2020-MulRIS}
P.~Wang, J.~Fang, X.~Yuan, Z.~Chen, and H.~Li, ``Intelligent reflecting
  surface-assisted millimeter wave communications: Joint active and passive
  precoding design,'' \emph{IEEE Transactions on Vehicular Technology},
  vol.~69, no.~12, pp. 14\,960--14\,973, 2020.

\bibitem{Yang2022-MulRISEE}
Z.~Yang, M.~Chen, W.~Saad, W.~Xu, M.~Shikh-Bahaei, H.~V. Poor, and S.~Cui,
  ``Energy-efficient wireless communications with distributed reconfigurable
  intelligent surfaces,'' \emph{IEEE Transactions on Wireless Communications},
  vol.~21, no.~1, pp. 665--679, 2022.

\bibitem{Jinkyu2024-RIS}
J.~Lee, H.~Seo, and W.~Choi, ``Computation-efficient reflection coefficient
  design for graphene-based {RIS} in wireless communications,'' \emph{IEEE
  Transactions on Vehicular Technology}, vol.~73, no.~3, pp. 3663--3677, 2024.

\bibitem{Gao2020-RISFCNN}
J.~Gao, C.~Zhong, X.~Chen, H.~Lin, and Z.~Zhang, ``Unsupervised learning for
  passive beamforming,'' \emph{IEEE Communications Letters}, vol.~24, no.~5,
  pp. 1052--1056, 2020.

\bibitem{HaAn2023-DoubleRIS}
H.~An~Le, T.~Van~Chien, V.~D. Nguyen, and W.~Choi, ``Double {RIS}-assisted
  {MIMO} systems over spatially correlated rician fading channels and finite
  scatterers,'' \emph{IEEE Transactions on Communications}, vol.~71, no.~8, pp.
  4941--4956, 2023.

\bibitem{WXU2022-RISDQNN}
W.~Xu, L.~Gan, and C.~Huang, ``A robust deep learning-based beamforming design
  for {RIS}-assisted multiuser {MISO} communications with practical
  constraints,'' \emph{IEEE Transactions on Cognitive Communications and
  Networking}, vol.~8, no.~2, pp. 694--706, 2022.

\bibitem{Zhong2022-STARRISHybridDRL}
R.~Zhong, Y.~Liu, X.~Mu, Y.~Chen, X.~Wang, and L.~Hanzo, ``Hybrid reinforcement
  learning for {STAR-RISs}: A coupled phase-shift model based beamformer,''
  \emph{IEEE Journal on Selected Areas in Communications}, vol.~40, no.~9, pp.
  2556--2569, 2022.

\bibitem{Shen2023-GNNWireless}
Y.~Shen, J.~Zhang, S.~H. Song, and K.~B. Letaief, ``Graph neural networks for
  wireless communications: From theory to practice,'' \emph{IEEE Transactions
  on Wireless Communications}, vol.~22, no.~5, pp. 3554--3569, 2023.

\bibitem{Eisen2020-GNNPowercontrol}
M.~Eisen and A.~Ribeiro, ``Optimal wireless resource allocation with random
  edge graph neural networks,'' \emph{IEEE Transactions on Signal Processing},
  vol.~68, pp. 2977--2991, 2020.

\bibitem{Shen2021-GNNPA}
Y.~Shen, Y.~Shi, J.~Zhang, and K.~B. Letaief, ``Graph neural networks for
  scalable radio resource management: Architecture design and theoretical
  analysis,'' \emph{IEEE Journal on Selected Areas in Communications}, vol.~39,
  no.~1, pp. 101--115, 2021.

\bibitem{Chowdhury2021-GNNWMMSE}
A.~Chowdhury, G.~Verma, C.~Rao, A.~Swami, and S.~Segarra, ``Unfolding {WMMSE}
  using graph neural networks for efficient power allocation,'' \emph{IEEE
  Transactions on Wireless Communications}, vol.~20, no.~9, pp. 6004--6017,
  2021.

\bibitem{Jiang2021-GNNRIS}
T.~Jiang, H.~V. Cheng, and W.~Yu, ``Learning to reflect and to beamform for
  intelligent reflecting surface with implicit channel estimation,'' \emph{IEEE
  Journal on Selected Areas in Communications}, vol.~39, no.~7, pp. 1931--1945,
  2021.

\bibitem{Lyu2024-RISGNN}
S.~Lyu, L.~Peng, and S.~Y. Chang, ``Investigating large-scale {RIS}-assisted
  wireless communications using {GNN},'' \emph{IEEE Transactions on Consumer
  Electronics}, vol.~70, no.~1, pp. 811--818, 2024.

\bibitem{Wang2023-RISGNNFL}
Z.~Wang, Y.~Zhou, Y.~Zou, Q.~An, Y.~Shi, and M.~Bennis, ``A graph neural
  network learning approach to optimize {RIS}-assisted federated learning,''
  \emph{IEEE Transactions on Wireless Communications}, vol.~22, no.~9, pp.
  6092--6106, 2023.

\bibitem{Liu2023-RISGNN}
M.~Liu, C.~Huang, M.~Di~Renzo, M.~Debbah, and C.~Yuen, ``Cooperative
  beamforming and {RISs} association for multi-{RISs} aided multi-users mmwave
  {MIMO} systems through graph neural networks,'' in \emph{ICC 2023 - IEEE
  International Conference on Communications}, 2023, pp. 4286--4291.

\bibitem{Guo2022-HetnetBF}
J.~Guo and C.~Yang, ``Learning power allocation for multi-cell-multi-user
  systems with heterogeneous graph neural networks,'' \emph{IEEE Transactions
  on Wireless Communications}, vol.~21, no.~2, pp. 884--897, 2022.

\bibitem{luo2008dynamic}
Z.-Q. Luo and S.~Zhang, ``Dynamic spectrum management: Complexity and
  duality,'' \emph{IEEE journal of selected topics in signal processing},
  vol.~2, no.~1, pp. 57--73, 2008.

\bibitem{Miguel1998-SecondoderCone}
M.~S. Lobo, L.~Vandenberghe, S.~Boyd, and H.~Lebret, ``Applications of
  second-order cone programming,'' \emph{Linear Algebra and its Applications},
  vol. 284, no.~1, pp. 193--228, 1998.

\bibitem{Chen2023-RISCE}
J.~Chen, Y.-C. Liang, H.~V. Cheng, and W.~Yu, ``Channel estimation for
  reconfigurable intelligent surface aided multi-user mmwave mimo systems,''
  \emph{IEEE Transactions on Wireless Communications}, vol.~22, no.~10, pp.
  6853--6869, 2023.

\bibitem{Zhang2021-PowerHetGNN}
X.~Zhang, H.~Zhao, J.~Xiong, X.~Liu, L.~Zhou, and J.~Wei, ``Scalable power
  control/beamforming in heterogeneous wireless networks with graph neural
  networks,'' in \emph{2021 IEEE Global Communications Conference (GLOBECOM)},
  2021, pp. 01--06.

\bibitem{Kurt1989-Universal}
K.~Hornik, M.~Stinchcombe, and H.~White, ``Multilayer feedforward networks are
  universal approximators,'' \emph{Neural Networks}, vol.~2, no.~5, pp.
  359--366, 1989.

\bibitem{Diederik2015-ADAM}
D.~P. Kingma and J.~Ba, ``Adam: {A} method for stochastic optimization,'' in
  \emph{3rd International Conference on Learning Representations, {ICLR} 2015,
  San Diego, CA, USA, May 7-9, 2015, Conference Track Proceedings}, Y.~Bengio
  and Y.~LeCun, Eds., 2015.

\bibitem{Eisen2019-MLResourceAllo}
M.~Eisen, C.~Zhang, L.~F.~O. Chamon, D.~D. Lee, and A.~Ribeiro, ``Learning
  optimal resource allocations in wireless systems,'' \emph{IEEE Transactions
  on Signal Processing}, vol.~67, no.~10, pp. 2775--2790, 2019.

\bibitem{He2022-GBLink}
S.~He, S.~Xiong, W.~Zhang, Y.~Yang, J.~Ren, and Y.~Huang, ``{GBLinks}:
  {GNN}-based beam selection and link activation for ultra-dense {D2D} {mmWave}
  networks,'' \emph{IEEE Transactions on Communications}, vol.~70, no.~5, pp.
  3451--3466, 2022.

\bibitem{HaAn2021-MLCE}
H.~A. Le, T.~Van~Chien, T.~H. Nguyen, H.~Choo, and V.~D. Nguyen, ``Machine
  learning-based {5G}-and-beyond channel estimation for {MIMO-OFDM}
  communication systems,'' \emph{Sensors}, vol.~21, no.~14, 2021.

\bibitem{SZhang2020-RISCapacity}
S.~Zhang and R.~Zhang, ``Capacity characterization for intelligent reflecting
  surface aided {MIMO} communication,'' \emph{IEEE Journal on Selected Areas in
  Communications}, vol.~38, no.~8, pp. 1823--1838, 2020.

\bibitem{pytorch}
A.~Paszke, S.~Gross, F.~Massa, A.~Lerer, J.~Bradbury \emph{et~al.}, ``Pytorch:
  An imperative style, high-performance deep learning library,'' in
  \emph{Advances in Neural Information Processing Systems 32}.\hskip 1em plus
  0.5em minus 0.4em\relax Curran Associates, Inc., 2019, pp. 8024--8035.

\bibitem{Adam-2014}
\BIBentryALTinterwordspacing
D.~P. Kingma and J.~Ba, ``Adam: A method for stochastic optimization,''
  \emph{CoRR}, vol. abs/1412.6980, 2014. [Online]. Available:
  \url{https://api.semanticscholar.org/CorpusID:6628106}
\BIBentrySTDinterwordspacing

\bibitem{HSONG2021-RISUnsuperved}
H.~Song, M.~Zhang, J.~Gao, and C.~Zhong, ``Unsupervised learning-based joint
  active and passive beamforming design for reconfigurable intelligent surfaces
  aided wireless networks,'' \emph{IEEE Communications Letters}, vol.~25,
  no.~3, pp. 892--896, 2021.

\bibitem{Scutari2017-SCA}
G.~Scutari, F.~Facchinei, and L.~Lampariello, ``Parallel and distributed
  methods for constrained nonconvex optimization—part i: Theory,'' \emph{IEEE
  Transactions on Signal Processing}, vol.~65, no.~8, pp. 1929--1944, 2017.

\end{thebibliography}
\end{document}